\documentclass[a4paper,11pt]{article}
\pdfoutput=1 

\usepackage{jcappub} 
\usepackage{natbib}
\usepackage{aas_macros}

\usepackage[T1]{fontenc} 
\usepackage{graphicx}
\usepackage{epsfig}
\usepackage{rotate}
\usepackage{amsmath,amssymb,amsfonts,mathrsfs}
\allowdisplaybreaks
\usepackage{bm}
\usepackage{xspace}
\usepackage{diagbox}
\usepackage{slashbox}
\usepackage{verbatim}
\usepackage{tablefootnote}
\usepackage{enumerate}
\usepackage{afterpage}
\usepackage{xcolor}
\usepackage{natbib}
\usepackage[nameinlink,capitalise]{cleveref}

\newcommand{\bk}{\bm{k}}
\newcommand{\bq}{\bm{q}}
\newcommand{\bx}{\bm{x}}
\newcommand{\fnl}{f_{\rm NL}}
\newcommand{\fnll}{f_{\mathrm{NL}}^{\mathrm{loc}}}
\newcommand{\fnle}{f_{\mathrm{NL}}^{\mathrm{equil}}}

\newcommand{\Mpc}{\ensuremath{\text{$h$/Mpc}}\xspace}
\newcommand*{\order} [1] {\ensuremath{\mathcal{O}(#1)}\xspace}
\renewcommand*\bar[1]{\overline{#1}}
\newcommand*{\veps} {\varepsilon}

\def\l{\big(}
\def\r{\big)}

\graphicspath{{./plots/}}

\title{Multi-tracer power spectra and bispectra: Formalism}

\author{Dionysios Karagiannis,$^{1}$ Roy Maartens,$^{1,2,3}$ 
Jos\'e Fonseca,$^{4,5,1}$ Stefano Camera,$^{6,7,8,1}$ Chris Clarkson$^{9,1}$
}
\affiliation{$^{1}$Department of Physics \& Astronomy, University of the Western Cape, Cape Town 7535, South Africa\\
$^{2}$Institute of Cosmology \& Gravitation, University of Portsmouth, Portsmouth PO1 3FX, United Kingdom\\
$^{3}$National Institute for Theoretical and Computational Sciences (NITheCS), Cape Town 7535, South Africa\\
$^4$Instituto de Astrofisica e Ciencias do Espaco, Universidade do Porto CAUP, 4150-762 Porto, Portugal\\
$^5$Departamento de F\'isica e Astronomia, Faculdade de Ci\^{e}ncias, Universidade do Porto, Rua do Campo Alegre 687, PT4169-007 Porto, Portugal\\
$^6$Dipartimento di Fisica, Universit\`a degli Studi di Torino, 10125 Torino, Italy\\
$^7$INFN -- Istituto Nazionale di Fisica Nucleare, Sezione di Torino, 10125 Torino, Italy\\
$^8$INAF -- Istituto Nazionale di Astrofisica, Osservatorio Astrofisico di Torino, 10025 Pino Torinese, Italy \\
$^9$Department of Physics \& Astronomy, Queen Mary University of London, London E1 4NS, United Kingdom
}

\emailAdd{dakaragian@gmail.com}

\abstract{
The power spectrum and bispectrum of dark matter tracers are key and complementary probes of the Universe. Next-generation surveys will deliver good measurements of the bispectrum, opening the door to improved cosmological constraints and the breaking of parameter degeneracies, from the combination of the power spectrum and bispectrum.  Multi-tracer power spectra have been used to suppress cosmic variance and mitigate the effects of nuisance parameters and systematics. We present a bispectrum multi-tracer   formalism that can be applied to next-generation survey data. Then we perform a simple Fisher analysis to illustrate qualitatively the improved precision on primordial non-Gaussianity that is expected to come from the bispectrum multi-tracer. In addition, we investigate the parametric dependence of conditional errors from multi-tracer power spectra and multi-tracer bispectra, on the differences between the biases and the number densities of two tracers. Our results suggest that optimal constraints arise from maximising the ratio of number densities, the difference between the linear biases, the difference between the quadratic  biases, and the difference  between the products $b_1\,b_\Phi$ for each tracer, where $b_\Phi$ is the bias for the primordial potential.
}

\begin{document}
\maketitle

\flushbottom

\section{Introduction}

Spectacular advances in observations and computations have transformed our understanding of the Universe. At the same time, this has brought new puzzles to the fore, such as the still unknown nature of dark matter and dark energy that are core to the current standard model, and the emergence of tensions between observations that appear to challenge the foundations of the standard model.  The next generation of large-scale structure surveys will  deliver major advances in precision cosmology, cast new light on current tensions and anomalies, and open new windows on the primordial and late-time Universe. 

Up to now, cosmological information has mainly been extracted via the galaxy power spectrum, but the galaxy bispectrum is set to play an increasingly important role, in combination with the power spectrum (see e.g.\ the recent work \cite{Novell-Masot:2023rli,Sugiyama:2023tes,Ivanov:2023qzb,Karagiannis:2022ylq,Philcox:2021kcw,Ivanov:2021kcd,MoradinezhadDizgah2020} and references therein). Recent measurements of the BOSS bispectrum \cite{Sugiyama:2023tes,Ivanov:2023qzb,Philcox:2021kcw}, including the first measurement of the bispectrum quadrupole \cite{Sugiyama:2018yzo}, prepare the way for next-generation surveys. These upcoming surveys are also predicted to detect the bispectrum dipole \cite{Clarkson:2018dwn,Maartens:2019yhx,Jolicoeur:2020eup} and other relativistic effects in the bispectrum \cite{Umeh:2016nuh,DiDio:2016gpd,Maartens:2020jzf,Noorikuhani:2022bwc}. Wide-angle effects in the galaxy bispectrum \cite{DiDio:2018unb,Garcia:2020per,Noorikuhani:2022bwc,Pardede:2023ddq} will also need to be taken into account.

The bispectrum is a direct probe of non-Gaussianity, and one of its key strengths lies in constraints on primordial non-Gaussianity (PNG). Unlike the power spectrum, it can constrain non-local types of PNG, while in combination with the power spectrum it can improve constraints on local PNG \cite{Karagiannis:2019jjx,Bharadwaj:2020wkc,Karagiannis:2020dpq,DAmico:2022gki,Coulton:2022rir,Jung:2022gfa}. In order to further improve the precision of power spectrum constraints, one can use a multi-tracer analysis, which suppresses cosmic variance and also reduces the effects of nuisance parameters and systematics (see e.g.\  \cite{Seljak:2008xr,McDonald:2008sh,Hamaus:2011dq,Abramo:2013awa}). A multi-tracer analysis is particularly effective for constraints on local PNG and other quantities that have a signal on very large scales, where cosmic variance is strongest (see e.g.\ \cite{Ferramacho:2014pua,Yamauchi:2014ioa,Alonso:2015sfa,Fonseca:2015laa,Fonseca:2016xvi,Gleyzes:2016tdh,Ballardini:2019wxj,Gomes:2019ejy,Viljoen:2021ypp,Barreira:2021ueb,Jolicoeur:2023tcu,Barreira:2023rxn,Sullivan:2023qjr}). 

This suggests the use of a multi-tracer bispectrum analysis, in combination with the corresponding multi-tracer power spectra. Here we develop for the first time a Fisher forecast multi-tracer formalism for the tree-level power spectra + bispectra combination.
Multi-tracer power spectra have been combined with the auto-bispectrum of one of the tracers by \cite{Barreira:2021ueb}. We extend such approaches to include all the bispectra. A multi-tracer forecast has been applied to bispectra alone (in real space) by \cite{Yamauchi2017}, on the basis of simplifying assumptions and without giving  the covariances or other details. We fill the gap by providing a systematic analysis in redshift-space,  including all clustering bias and noise parameters at tree level and giving explicit forms for the covariances. 

In this first paper, we do not perform detailed forecasts, which will be presented in future work. However, we do show qualitatively the improvements that are potentially delivered by the full power spectrum and bispectrum multi-tracer of 2 tracers, by investigating the parametric behaviour of the conditional errors on PNG. To quantify the potential of this method we consider a comparison between the forecasts generated by the multi-tracer and each single-tracer, over a single volume and with the same redshift range. This is done in order to investigate the effect of combining the samples or not on primordial non-Gaussianity, without the effect of survey specifications for the different tracers, e.g. new volumes and different redshift overlap. To begin with, we do not include wide-angle or relativistic effects. 

\section{Single-tracer power spectrum and bispectrum}\label{sec:stpb}

The power spectrum of the Bardeen gauge-invariant primordial gravitational potential is defined in Fourier space by
\begin{equation}
 {\langle \Phi(\bk)\,\Phi(\bk')\rangle}=(2\,\pi)^3\,\delta_{\rm D}(\bk+\bk')\,P_\Phi(k)\;,  
\end{equation}
where $P_\Phi(k)$ is directly related to the power spectrum of the primordial curvature perturbations $\zeta$ (during the matter-dominated era, $\Phi=3\,\zeta/5$), which are generated during inflation.  In the case of the standard single-field slow-roll inflationary scenario, they have a nearly perfect Gaussian distribution, which means that they can be adequately characterised by their power spectrum. The primordial perturbations $\Phi$ are in turn related to the linear dark matter over-density field through the Poisson equation, ${\delta}  (\bk,z)=M(k,z)\,\Phi(\bk)$, where
 \begin{align}
 {P}(k,z)&=M(k,z)^2\,P_\Phi(k)\;,\nonumber\\
 M(k,z)&=\frac{2\,D(z)}{3\,\Omega_{{\rm m},0} \,H_0^2\, {g_{\rm dec}}}\,T(k)\, k^2\;.\label{eq:poisM}
 \end{align}
 Here $D(z)$ is the linear growth factor (since the linear fluid equations generate a linearly evolved matter density field), normalised to unity at $z=0$, and $T(k)$ is the matter transfer function normalized to unity at large scales $k \rightarrow 0$. {The factor $g_{\rm dec}$ is the Bardeen potential growth factor at decoupling, which ensures that $\fnl$ is in the CMB convention \cite{Camera:2014bwa,Desjacques2016}.}
 The linear power spectrum is computed with the numerical Boltzmann code \texttt{CAMB} \cite{CAMB}. 
 
 Deviations from the standard inflationary model will produce a violation of Gaussian initial conditions \cite{Komatsu2009}. PNG generates nonzero higher-order correlators in the primordial curvature perturbations, starting with the bispectrum, i.e.\ the Fourier transform of the three-point function,
 \begin{equation}
\langle\Phi(\bk_1)\,\Phi(\bk_2)\,\Phi(\bk_3)\rangle=(2\,\pi)^3\,\delta_{\rm D}(\bk_1+\bk_2+\bk_3)\,B_{\Phi}(k_1,k_2,k_3)\;.
 \end{equation}
 Higher-order correlators in the primordial density field  are a product of nonlinear interactions during the inflationary or reheating stage. The primordial bispectrum arising from such interactions
is characterized by a dimensionless amplitude parameter
$\fnl$, giving the strength of the PNG signal, and  a shape function $F$, i.e.\ $B_{\Phi}(k_1,k_2,k_3)=\fnl\,F(k_1, k_2, k_3)$. $F$ describes the dependence of the bispectrum on various triangles, where different inflationary models generate signal that peaks at distinct triangle configurations. 
By \cref{eq:poisM}, the leading order PNG contribution to the matter density bispectrum is
 \begin{equation} \label{eq:bisng}
  B_{\rm PNG}(k_1,k_2,k_3,z)=M(k_1,z)\,M(k_2,z)\,M(k_3,z)\,B_\Phi(k_1,k_2,k_3) \;.
 \end{equation}

The matter bispectrum has additional  contributions from nonlinearity induced by gravity, even at zeroth order. In other words, the dark matter density field is intrinsically non-Gaussian, even in the absence of PNG. Therefore, in order to disentangle the PNG information from late-time nonlinearity, an adequate description for the latter is needed. Here we model nonlinearity in the framework of Standard Perturbation Theory (SPT) (see e.g.\ \cite{Bernardeau2002} for a review). 
In addition to the nonlinear nature of the dark matter density field, the hierarchical formation of halos and the way they are populated by luminous matter induces non-Gaussianity in the distribution of observable structures. Hence, in order to extract the PNG signal from large-scale structure, we need the bias relationship between the observed tracers and the matter field.

We follow the perturbative approach of \cite{Assassi2014,Senatore2014,Mirbabayi2014}, in which the halo over-density field $\delta_h$ is described as a function of all possible local gravitational observables, which are introduced in the expansion in the form of renormalised operators. A complete Eulerian bias expansion can be built from the tensor $\partial_i\partial_j\Phi$, which 
contains the trace $\nabla^2\Phi\propto\delta $ and the trace-free  tidal field $s_{ij}=\l\partial_{i}\partial_{j}-\delta_{ij}\,\nabla^2/3\r\,\nabla^{-2} \delta $. Note that the term `halo' here can be replaced by a general matter tracer. 

We consider up to second-order terms in the expansion, which are sufficient for the spatial scales considered here, i.e.\ much larger scales than those involved in halo formation. For Gaussian initial conditions, the Eulerian halo density contrast is  \cite{Assassi2014,Senatore2014,Mirbabayi2014,Desjacques2016}:   
\begin{equation}\label{eq:deltaG}
   {\delta^h}(\bx,\tau)= b_1(\tau)\,\delta(\bx,\tau) +\veps(\bx,\tau)+\frac{b_2(\tau)}{2}\,\delta(\bx,\tau)^2 + \frac{b_{s}(\tau)}{2}\,s(\bx,\tau)^2+\veps_{\delta}(\bx,\tau)\,\delta(\bx,\tau)\;.
  \end{equation}
  Here $\tau$ is  conformal time, $\bx$ are  spatial comoving Eulerian coordinates, $s^2=s_{ij}\,s^{ij}$
  is the simplest scalar that can be formed from the tidal field,  
  $\veps$ is the leading stochastic field \cite{Dekel1998,Taruya1998,Matsubara1999} and $\veps_{\delta}$ is the stochastic field associated with the linear bias. These fields take into account the stochastic relation between the tracer density and any large-scale field. The second-order tidal field bias coefficient, following the convention in \cite{Baldauf2012}, is given by $b_{s}=-4\,(b_1-1)/7$. 
  
  In the presence of PNG there is a scale-dependent correction to the linear bias $b_1$, especially in the case of local PNG \cite{Dalal2008,Slosar2008,Matarrese2008,Verde2009,Afshordi2008,Desjacques2010}, since the local PNG bispectrum peaks in squeezed triangles. (A similar scale-dependent bias correction can be derived for any general nonlocal quadratic non-Gaussianity template \cite{Schmidt2010,Scoccimarro2011,Desjacques2011b,Schmidt2013}.) Then the non-Gaussian bias terms, linear in $\fnl$, are \cite{Assassi2015}:
  \begin{equation}\label{eq:deltaNG}
   {\delta^h_{\rm PNG}}(\bx,\tau)=b_{\Psi}(\tau)\,\Psi(\bq)+b_{\Psi\delta}(\tau)\,\Psi(\bq)\,\delta(\bx,\tau)+\veps_{\Psi}(\bx,\tau)\,\Psi(\bq)\;,
\end{equation}
where $\bq$ are the spatial coordinates in the Lagrangian frame and $\Psi$ is a nonlocal transformation of $\Phi$: 
\begin{align}\label{psiphi}
\Psi(\bq)=\int \frac{{\rm d}^3\bk}{(2\,\pi)^3}\, k^\alpha \, \Phi(\bk)\, {\rm e}^{{\rm i}\,\bq \cdot\bk}\;.
\end{align}
{The parameter $\alpha=2,1,0$ for equilateral, orthogonal and local shapes. Note that in the local case, $\Psi\rightarrow\Phi$.}
The stochastic counterpart of $\Psi$ is  $\varepsilon_\Psi$. The bias coefficients in \cref{eq:deltaG,,eq:deltaNG} can be derived by using the peak-background split  argument (see e.g.\  \cite{Karagiannis:2020dpq}). 
  
{The single-tracer power spectrum and  bispectrum in  redshift space are}
     \begin{align}
{P^h}(\bk,z)&={\cal D}_P(\bk,z)\,Z_1(\bk,z)^2\,P(k,z)+P_{\veps}(z) \label{eq:Pgs},\\ 
{B^h}(\bk_1,\bk_2,\bk_3,z)&= {\cal D}_B(\bk_1,\bk_2,\bk_3,z) \,\Big[2\,Z_1(\bk_1,z)\,Z_1(\bk_2,z)\,Z_2(\bk_1,\bk_2,z)\,P(k_1,z)\,P(k_2,z)\nonumber \\
   &~~+2\,\text{perm}\Big] +2\,P_{\veps\veps_{\delta}}(z)\,\Big[Z_1(\bk_1,z)\,P(k_1,z)+2\,\text{perm}\Big]+B_{\veps}(z)\;. \label{eq:Bgs} 
  \end{align}
At tree level, the first- and second-order kernels $Z_{1,2}$ include redshift space distortions (RSD) \cite{Sargent:1977,Kaiser1987,Hamilton1998} and PNG. In order to deal with redshift errors and nonlinear RSD, we include the phenomenological factors ${\cal D}_{P,B}$. Explicit expressions are in \cref{app:RSD_kernels}.
In the case of power spectra, an alternative to the phenomenological factor ${\cal D}_{P}$ is to use the 1-loop EFT model, as in the multi-tracer power spectrum analysis of \cite{Mergulhao:2021kip}.
The stochastic terms  $P_{\veps},\;P_{\veps\veps_{\delta}},\;B_{\veps}$ are generated by  stochastic bias and their fiducial values are taken to be those predicted by Poisson statistics \cite{Schmidt2015}:  
  \begin{equation}\label{stoch-st}
   P_{\veps}=\frac{1}{{\bar{n}_h}}\;,~~P_{\veps\veps_{\delta}}=\frac{b_1}{2\,\bar{n}_h}\;,~~B_{\veps}=\frac{1}{\bar{n}^2_h}\;.
  \end{equation}

\section{Multi-tracer power spectra and bispectra}

In the multi-tracer framework (e.g.\ \cite{Seljak:2008xr,McDonald:2008sh,Hamaus:2011dq,Abramo:2013awa,Ferramacho:2014pua,Yamauchi:2014ioa,Alonso:2015sfa,Fonseca:2015laa,Fonseca:2016xvi,Gleyzes:2016tdh,Ballardini:2019wxj,Gomes:2019ejy,Viljoen:2021ypp,Barreira:2021ueb,Jolicoeur:2023tcu,Barreira:2023rxn,Sullivan:2023qjr}) we consider the combined information of all correlations of different tracer samples in a given redshift bin and sky patch. The  multi-tracer power spectra and bispectra are 
\begin{align}
 \langle \delta^I(\bk_1)\,\delta^J(\bk_2)\rangle &= (2\,\pi)^3\, \delta_{\rm D}(\bk_1+\bk_2)\,P^{IJ}(\bk_1),\\
 \langle \delta^I(\bk_1)\,\delta^J(\bk_2)\,\delta^K(\bk_3)\rangle &= (2\,\pi)^3 \,\delta_{\rm D}(\bk_1+\bk_2+\bk_3)\,B^{IJK}(\bk_1,\bk_2,\bk_3)\;,
\end{align}
where here and elsewhere we omit the $z$-dependence for brevity. The tracer indices $I,J,K$ range over the tracer labels: $I,J,K=t,t',t'',\cdots$.
 Note that the cross-power spectra are symmetric under permutations of the tracers, while the cross-bispectra require symmetrisation in order to avoid double counting of information: 
\begin{align} 
P^{(IJ)} & \equiv P^{IJ},
\\ \label{symbis}
{B}^{(IJK)}&=\frac{1}{6}\,\left({B}^{IJK}+{B}^{KJI}+{B}^{KIJ}+{B}^{JIK}+{B}^{JKI}+{B}^{IKJ}\right)\;.
\end{align}
{We will also use the short-hand notation}
\begin{align}
P^I \equiv P^{II} ,\quad B^I \equiv 
B^{III} \;,
\end{align}
where convenient.
 
 We define the estimators:
\begin{align}\label{hatp}
\hat{P}^{IJ}(\bk)&=\frac{1}{V_s\,V_k}\,\sum_{\bq \in k}\delta^I(\bq)\,\delta^J(-\bq)\;, \\
\hat{B}^{(IJK)}(\bk_1,\bk_2,\bk_3)&=\frac{1}{V_s\,V_{123}}\sum_{\bq_1\in k_1}\sum_{\bq_2\in k_2}\sum_{\bq_3\in k_3}
\delta_{\rm K}(\bq_{123})\,\delta^I(\bq_1)\,\delta^J(\bq_2)\,\delta^K(\bq_3)\;,
\label{hatb}
\end{align}
where $\bq_{123}\equiv\bq_1+\bq_2+\bq_3$. For the power spectra the sum runs over all wavenumbers $\bq$ in the spherical shell of radius $k$ and width $\Delta k$. For the bispectra, the Kronecker delta symbol ensures the formation of `fundamental triangles' with sides $\bq_i$, satisfying the condition $\bq_{123}=0$, that fall in the `triangular bin' defined by the triplet of bin centres ($k_1$, $k_2$, $k_3$) and width $\Delta k$. 
The volume in Fourier space of the power spectra shells and the bispectra  fundamental triangle bins is, in the thin shell limit \cite{Sefusatti2006},
\begin{align}\label{eq:vk}
V_k =4\pi\, k^2\,\Delta k\,, \quad 
V_{123}\equiv
\sum_{\bq_1\in k_1}\sum_{\bq_2\in k_2}\sum_{\bq_3\in k_3}
\delta_{\rm K}(\bq_{123})={8\pi^2}\,k_1\,k_2\,k_3\,\Delta k_1\,\Delta k_2\,\Delta k_3\,. 
\end{align}  
The  estimators in \cref{hatp,,hatb} 
in the case of one sample ($I=J=K$) reduce to the well known auto-correlation results.

\subsection*{The two-tracer case}\label{sec:2tr_case_model}

From now on we consider the case of two tracers, $t$ and $t'$, so that $I,J,K=t,t'$. The data vector of the power spectra is
\setlength{\abovedisplayskip}{9pt}
\setlength{\belowdisplayskip}{9pt}
\begin{equation}\label{eq:PSdata}
\bm{D_P}=\Big[P^{tt},P^{tt'},P^{t't'}\Big]\;.
\end{equation}
The expected values of the spectra are:
\begin{equation}
\langle \hat{P}^{IJ}\rangle\equiv P^{IJ}(\bk,z)={\cal D}_P^{IJ}(\bk,z)\,Z_1^I(\bk,z)\,Z_1^J(\bk,z)\,P(k,z)+\delta^{IJ}\,P^{I}_{\veps}(z)\,,
\label{eq:PS_MT}    
\end{equation}
%
where $Z_1^I$ and ${\cal D}_P^{IJ}$  are given in \cref{app:RSD_kernels}. 

The data vector of the bispectrum is 
\begin{equation}\label{eq:BSdata}
\bm{D_B}=\Big[B^{ttt},B^{(ttt')},B^{(tt't')},B^{t't't'}\Big]\;.
\end{equation}
In a tree-level description, we consider terms up to second order PT theory,  RSD and bias, while keeping only linear $\fnl$ terms in the perturbative expansion. Then the expected values of the estimator for each bispectrum data vector entry are:
\begin{align}
\langle \hat{B}^{ttt}\rangle\equiv B^{ttt}(\bk_1,\bk_2,\bk_3)&= {\cal D}_B^{ttt}\,\bigg[Z_1^t (\bk_1)\,Z_1^t (\bk_2)\,Z_1^t (\bk_3)\,B_{\rm PNG}(k_1,k_2,k_3) \nonumber \\ 
   &+\Big\{2\,Z_1^t (\bk_1)\,Z_1^t (\bk_2)\,Z_2^t (\bk_1,\bk_2)\,P(k_1)\,P(k_2)+2\, \bk\text{-perm}\Big\}\bigg] \nonumber \\
   &+2\,P^t _{\veps\veps_{\delta}}\,\Big[Z_1^t (\bk_1)\,P(k_1)+2\,\bk\text{-perm}\Big]+B^t _{\veps}\;, \label{eq:BS_MT1} \\
  \langle \hat{B}^{(ttt')}\rangle\equiv B^{(ttt')}(\bk_1,\bk_2,\bk_3)&= \frac{1}{3}\,\bigg[\Big\{{\cal D}_B^{ttt'}\,Z_1^t (\bk_1)\,Z_1^t (\bk_2)\,Z_1^{t'} (\bk_3)+ {\cal D}_B^{tt't}\,Z_1^t (\bk_1)\,Z_1^{t'} (\bk_2)\,Z_1^t (\bk_3) \nonumber \\
&+ {\cal D}_B^{t'tt}Z_1^{t'} (\bk_1)\,Z_1^t (\bk_2)\,Z_1^t (\bk_3)\Big\}\,B_{\rm PNG}(k_1,k_2,k_3) \nonumber \\ 
   &+2\,\Big\{\big[{\cal D}_B^{tt't}\,Z_1^t (\bk_1)\,Z_1^{t'} (\bk_2)+{\cal D}_B^{t'tt}\,Z_1^{t'} (\bk_1)\,Z_1^t (\bk_2)\big]\,Z_2^t (\bk_1,\bk_2) \nonumber \\
  &+{\cal D}_B^{ttt'}\,Z_1^t (\bk_1)\,Z_1^t (\bk_2)\,Z_2^{t'} (\bk_1,\bk_2) \Big\}\,P(k_1)\,P(k_2)+2\,\bk\text{-perm}\bigg]\nonumber \\
  & +\frac{1}{3}\,P^t _{\veps\veps_\delta}\,\Big[Z^{t'} (\bk_1)\,P(k_1)+2\,\bk\text{-perm}\Big]\;, \label{eq:BS_MT112} \\
\langle \hat{B}^{(tt't')}\rangle\equiv B^{(tt't')}(\bk_1,\bk_2,\bk_3)&= \frac{1}{3}\,\bigg[ \Big\{{\cal D}_B^{tt't'}\,Z_1^t (\bk_1)\,Z_1^{t'} (\bk_2)\,Z_1^{t'} (\bk_3)+ {\cal D}_B^{t't't}\,Z_1^{t'} (\bk_1)\,Z_1^{t'} (\bk_2)\,Z_1^t (\bk_3) \nonumber \\
&+ {\cal D}_B^{t'tt'}\,Z_1^{t'} (\bk_1)\,Z_1^t (\bk_2)\,Z_1^{t'} (\bk_3)\Big\}\,B_{\rm PNG}(k_1,k_2,k_3) \nonumber \\ 
   &+2\,\Big\{\big[{\cal D}_B^{tt't'}\,Z_1^t (\bk_1)\,Z_1^{t'} (\bk_2)+{\cal D}_B^{t'tt'}\,Z_1^{t'} (\bk_1)\,Z_1^t (\bk_2)\big]\,Z_2^{t'} (\bk_1,\bk_2) \nonumber \\
  &+{\cal D}_B^{t't't}\,Z_1^{t'} (\bk_1)\,Z_1^{t'} (\bk_2)\,Z_2^t (\bk_1,\bk_2) \Big\}P(k_1)\,P(k_2)+2\,\bk\text{-perm}\bigg]\nonumber \\
  & +\frac{1}{3}\,{P^{t'} _{\veps\veps_\delta}}\,\Big[Z^t (\bk_1)\,P(k_1)+2\,\bk\text{-perm}\Big]\;, \label{eq:BS_MT122} \\
\langle \hat{B}^{t't't'}\rangle\equiv B^{t't't'}(\bk_1,\bk_2,\bk_3)&= {\cal D}_B^{t't't'}\,\bigg[Z_1^{t'} (\bk_1)\,Z_1^{t'} (\bk_2)\,Z_1^{t'} (\bk_3)\,B_{\rm PNG}(k_1,k_2,k_3) \nonumber \\ 
   &+\Big\{2\,Z_1^{t'} (\bk_1)\,Z_1^{t'} (\bk_2)\,Z_2^{t'}(\bk_1,\bk_2)\,P(k_1)\,P(k_2)+2\,\bk\text{-perm}\Big\}\bigg] \nonumber \\
   &+2\,P^{t'} _{\veps\veps_{\delta}}\,\Big[Z_1^{t'} (\bk_1)\,P(k_1)+ 
   2\,\bk\text{-perm}\Big]+B^{t'} _{\veps}\;, \label{eq:BS_MT2}
\end{align}
where $Z_2^I$  and ${\cal D}_B^{IJK}$ are given in \cref{app:RSD_kernels}. Note that we omitted the redshift dependence in all terms,  and also the $k$-dependence in ${\cal D}_B^{IJK}$. 
The  stochastic terms have fiducial values:
    \begin{equation}\label{eq:poisson_fid}
   P_{\veps}^I=\frac{1}{\bar{n}_I}\;,~~P_{\veps\veps_{\delta}}^I=\frac{\;b_1^{I}}{2\,\bar{n}_I}\;,~~ B_{\veps}^I=\frac{1}{\bar{n}_I^2}\;.
  \end{equation}
We assume that the cross shot noise of the power spectra and bispectra is zero:
\begin{equation}
P^{IJ}_\veps=P_{\veps\veps_{\delta}}^{IJ}= 0 ~~(I\neq J)\;,\quad
B^{IJK}_\veps=0 ~~(I,\,J,\,K~\mbox{unequal})\;.   \end{equation}
  


The extension of the power spectrum and bispectrum multi-tracers from 2 to 3 tracers is given in \cref{app:3t}. 

Note that the multi-tracer tree-level expressions listed above are for a general form of the redshift-space kernels $Z_1^I$ and $Z_2^I$. This means that additional effects that modify these kernels can be introduced fairly easily in the formalism. For example, we can incorporate local relativistic effects by replacing the kernels given in \cref{app:RSD_kernels} with those given in~\cite{Maartens:2020jzf}.

  \section{Multi-tracer covariance}\label{sec:covariance}


The multi-tracer combination of power spectra and bispectra for two tracers has data vector 
\begin{equation}
\bm{D_{PB}}=\Big[P^{tt},P^{tt'},P^{t't'},B^{ttt},B^{(ttt')},B^{(tt't')},B^{t't't'} \Big]\;.
\end{equation}
The covariance of the data vector contains four blocks, coming from the covariance of the power spectra, the bispectra and the power spectra -- bispectra covariance. In  matrix form:
\begin{equation}\label{eq:comb_COV}
\bm{\mathrm{Cov}_{PB}}=
\begin{bmatrix}
\mathrm{Cov}(\bm{P}_i, \bm P_j)~ & & \mathrm{Cov}(\bm{P}_i, \bm B_{{ b}}) \\
 & & \\
\mathrm{Cov}(\bm{P}_i, \bm B_{{ b}})^{\sf T} 
& & \mathrm{Cov}(\bm{B}_{{ a}}, \bm B_{{ b}})
\end{bmatrix}\;,
\end{equation}  
where 
\begin{align}
\bm{P}_i=\bm{P}(\bk_i)\;,\quad  
\bm{B}_{{ a}}=\bm{B}(\bk_{a_1}, \bk_{a_2}, \bk_{a_3})\;.
\end{align}

We assume that the covariance between two power spectra is Gaussian, i.e.\ the off-diagonal contributions are neglected. Then the cross-sample covariance of the power spectra is a symmetric 3$\times$3 diagonal block matrix:
  \begin{align}\label{eq:PS_cov}
  \mathrm{Cov}(\bm{P}_i,\bm{P}_j)&= \begin{pmatrix}
    \mathrm{Cov}\l P_i^{tt}, P_j^{tt}\r & \mathrm{Cov}\l P_i^{tt}, P_j^{tt'}\r &  \mathrm{Cov}\l P_i^{tt}, P_j^{t't'}\r\\
    &&\\
     & \mathrm{Cov}\l P_i^{tt'}, P_j^{tt'}\r & \mathrm{Cov}\l P_i^{tt'}, P_j^{t't'}\r \\
    &&\\
     & & \mathrm{Cov}\l P_i^{t't'}, P_j^{t't'}\r \\
    \end{pmatrix}\\ \nonumber  
  &=2\,\frac{(2\,\pi)^3}{V_s\,V_k}\,\delta_{ij}  
  \begin{pmatrix}
    \big[P^{tt}(\bk_i)\big]^2 & P^{tt}(\bk_i)\,P^{tt'}(\bk_i) & \big[P^{tt'}(\bk_i)\big]^2\\
    &&\\
     & \frac{1}{2}\,\big\{P^{tt}(\bk_i)\,P^{t't'}(\bk_i)+\big[P^{tt'}(\bk_i)\big]^2\big\} & P^{t't'}(\bk_i)\,P^{tt'}(\bk_i) \\
    &&\\
    &  & \big[P^{t't'}(\bk_i)\big]^2
    \end{pmatrix}\;,
  \end{align}
where 
each sub-block is a diagonal matrix  in $\bk_i$.
  
The covariance of the symmetrised bispectra is a {symmetric} 4$\times$4 block matrix:
    \begin{align}\label{eq:BS_cov}
  & \mathrm{Cov}(\bm{B}_{{ a}},\bm{B}_{{ b}}) =\frac{(2\,\pi)^6}{3\,V_s\,V_{123}}\,s_{123}\,\delta_{ab}\times \\ \nonumber
    &\begin{pmatrix}
    \mathrm{Cov}\l B_{{ a}}^{ttt}, B_{{ b}}^{ttt}\r & \mathrm{Cov}\l B_{{ a}}^{ttt}, B_{{ b}}^{(ttt')}\r &  \mathrm{Cov}\l B_{{ a}}^{ttt}, B_{{ b}}^{(tt't')}\r & \mathrm{Cov}\l B_{{ a}}^{ttt}, B_{{ b}}^{t't't'}\r\\
    &&&\\
    & \mathrm{Cov}\l B_{{ a}}^{(ttt')}, B_{{ b}}^{(ttt')}\r & \mathrm{Cov}\l B_{{ a}}^{(ttt')}, B_{{ b}}^{(tt't')}\r & \mathrm{Cov}\l B_{{ a}}^{(ttt')}, B_{{ b}}^{t't't'}\r \\
    &&&\\
     &  & \mathrm{Cov}\l B_{{ a}}^{(tt't')}, B_{{ b}}^{(tt't')}\r & \mathrm{Cov}\l B_{{ a}}^{(tt't')}, B_{{ b}}^{t't't'}\r \\
    &&&\\
     &  &  &  \mathrm{Cov}\l B_{{ a}}^{t't't'}, B_{{ b}}^{t't't'}\r
    \end{pmatrix}\;,
  \end{align}
where $s_{123}=6,2,1$ for equilateral, isosceles and scalene triangles respectively.  Each block is an $N_T\times N_T$ matrix, where $N_T$ is the total number of triangles formed. The index ${{ a}}$ represents the triangle triplet $(\bk_{a_1}, \bk_{a_2}, \bk_{a_3})$. The Kronecker delta indicates that each sub-block of the bispectrum-bispectrum covariance is a diagonal matrix, i.e.\ we only consider the Gaussian contribution.

These sub-block matrices  are:
\begin{align} \label{eq:blockBB}
\mathrm{Cov}\l B_a^{ttt}, B_b^{ttt}\r &= 3\,P^{tt}_{a_1}\,P^{tt}_{a_2}\,P^{tt}_{a_3}, \nonumber \\
\mathrm{Cov}\l B_a^{ttt}, B_b^{(ttt')}\r&= P^{tt}_{a_1}\,P^{tt}_{a_2}\,P^{tt'}_{a_3} + \text{2\,perm.}\;, \nonumber \\
\mathrm{Cov}\l B_a^{ttt}, B_b^{(tt't')}\r &= P^{tt}_{a_1}\,P^{tt'}_{a_2}\,P^{tt'}_{a_3} + \text{2\,perm.}\;, \nonumber \\
\mathrm{Cov}\l B_a^{ttt}, B_b^{t't't'}\r &= 3\,P^{tt'}_{a_1}\,P^{tt'}_{a_2}\,P^{tt'}_{a_3},  \nonumber \\
\mathrm{Cov}\l B_a^{(ttt')}, B_b^{(ttt')}\r &= \frac{1}{3}\,\l P^{tt}_{a_1}\,P^{tt}_{a_2}\,P^{t't'}_{a_3}+2\,P^{tt}_{a_1}\,P^{tt'}_{a_2}\,P^{tt'}_{a_3}\r + \text{2\,perm.}\;, \nonumber \\
\mathrm{Cov}\l B_a^{(ttt')}, B_b^{(tt't')}\r &= \frac{1}{3}\,\l P^{tt'}_{a_1}\,P^{tt'}_{a_2}\,P^{tt'}_{a_3}+P^{tt}_{a_1}\,P^{tt'}_{a_2}\,P^{t't'}_{a_3}+P^{t't'}_{a_1}\,P^{tt'}_{a_2}\,P^{tt}_{a_3}\r + \text{2\,perm.}\;, \nonumber \\
\mathrm{Cov}\l B_a^{(ttt')}, B_b^{t't't'}\r &= P^{tt'}_{a_1}\,P^{tt'}_{a_2}\,P^{t't'}_{a_3}+ \text{2\,perm.}\;,  \nonumber \\
\mathrm{Cov}\l B_a^{(tt't')}, B_b^{(tt't')}\r &= \frac{1}{3}\,\l P^{tt}_{a_1}\,P^{t't'}_{a_2}\,P^{t't'}_{a_3}+2\,P^{tt'}_{a_1}\,P^{tt'}_{a_2}\,P^{t't'}_{a_3}\r + \text{2\,perm.}\;, \nonumber \\
\mathrm{Cov}\l B_a^{(tt't')}, B_b^{t't't'}\r &= P^{tt'}_{a_1}\,P^{t't'}_{a_2}\,P^{t't'}_{a_3} + \text{2\,perm.}\;, \nonumber \\
\mathrm{Cov}\l B_a^{t't't'}, B_b^{t't't'}\r &= 3\,P^{t't'}_{a_1}\,P^{t't'}_{a_2}\,P^{t't'}_{a_3}\;.
\end{align}

Finally, the cross-covariance between the power spectra and bispectra is itself a 3$\times$4 block matrix:    

\vspace{2em}
\begin{align}\label{eq:CovPB}
  &\mathrm{Cov}(\bm{P}_i,\bm{B}_b) =\frac{(2\,\pi)^3}{V_s\,V_k}\\ \times \nonumber 
    &\begin{pmatrix}
    \mathrm{Cov}\l P_i^{tt}, B_b^{ttt}\r & \mathrm{Cov}\l P_i^{tt}, B_b^{(ttt')}\r &  \mathrm{Cov}\l P_i^{tt}, B_b^{(tt't')}\r & \mathrm{Cov}\l P_i^{tt}, B_b^{t't't'}\r\\
    &&&\\
    \mathrm{Cov}\l P_i^{tt'}, B_b^{ttt}\r & \mathrm{Cov}\l P_i^{tt'}, B_b^{(ttt')}\r & \mathrm{Cov}\l P_i^{tt'}, B_b^{(tt't')}\r & \mathrm{Cov}\l P_i^{tt'}, B_b^{t't't'}\r \\
    &&& \\
    \mathrm{Cov}\l P_i^{t't'}, B_b^{ttt}\r & \mathrm{Cov}\l P_i^{t't'}, B_b^{(ttt')}\r & \mathrm{Cov}\l P_i^{t't'}, B_b^{(tt't')}\r & \mathrm{Cov}\l P_i^{t't'}, B_b^{t't't'}\r 
    \end{pmatrix}\;,
  \end{align}
where 
\begin{align} \label{eq:blockPB}
\mathrm{Cov}\l P_i^{tt}, B_b^{ttt}\r &= {2\,\l\delta_{ib_1}+\delta_{ib_2}+\delta_{ib_3}\r \,P^{tt}({\bk}_i)\,B^{ttt}(\bk_{b_1},\bk_{b_2},\bk_{b_3})}\;, \nonumber \\
\mathrm{Cov}\l P_i^{tt}, B_b^{(ttt')}\r &= \frac{2}{3}\,\Big[\l\delta_{ib_1}+\delta_{ib_2}+\delta_{ib_3}\r\, P^{tt'}(\bk_i)\,B^{ttt}(\bk_{b_1},\bk_{b_2},\bk_{b_3})
\notag \\
&~~~+P^{tt}(\bk_i)B_{2\delta}^{ttt'}(\bk_{b_1},\bk_{b_2},\bk_{b_3})\Big]\;, \nonumber \\
\mathrm{Cov}\l P_i^{tt}, B_b^{(tt't')}\r &=\frac{2}{3}\,\Big[P^{tt}(\bk_i)\,B_{1\delta}^{tt't'}(\bk_{b_1},\bk_{b_2},\bk_{b_3})+P^{tt'}(\bk_i)\,B_{2\delta}^{ttt'}(\bk_{b_1},\bk_{b_2},\bk_{b_3})\Big]\;, \nonumber \\
\mathrm{Cov}\l P_i^{tt}, B_b^{t't't'}\r &= 2\,P^{tt'}(\bk_i)\,B_{1\delta}^{tt't'}(\bk_{b_1},\bk_{b_2},\bk_{b_3})\;,  \nonumber \\
\mathrm{Cov}\l P_i^{tt'}, B_b^{ttt}\r &=\big(\delta_{ib_1}+\delta_{ib_2}+\delta_{ib_3}\big)\,P^{tt'}(\bk_i)\,B^{ttt}(\bk_{b_1},\bk_{b_2},\bk_{b_3})+P^{tt}(\bk_i)\,B_{1\delta}^{ttt'}(\bk_{b_1},\bk_{b_2},\bk_{b_3})\;, \nonumber \\
\mathrm{Cov}\l P_i^{tt'}, B_b^{(ttt')}\r &=\frac{1}{3}\,\bigg\{ \l\delta_{ib_1}+\delta_{ib_2}+\delta_{ib_3}\r\,\Big[P^{tt'}(\bk_i)\,B^{(ttt')}(\bk_{b_1},\bk_{b_2},\bk_{b_3})
\notag \\
&~~~+P^{t't'}(\bk_i)\,B^{ttt}(\bk_{b_1},\bk_{b_2},\bk_{b_3})\Big] +P^{tt}(\bk_i)\,B_{2\delta}^{tt't'}(\bk_{b_1},\bk_{b_2},\bk_{b_3})\bigg\}\;, \nonumber \\
\mathrm{Cov}\l P_i^{tt'}, B_b^{(tt't')}\r &= \frac{1}{3} \,\bigg\{ \l\delta_{ib_1}+\delta_{ib_2}+\delta_{ib_3}\r \,\Big[P^{tt'}(\bk_i)\,B^{(tt't')}(\bk_{b_1},\bk_{b_2},\bk_{b_3})
\notag \\
&~~~+P^{t't'}(\bk_i)\,B^{ttt}(\bk_{b_1},\bk_{b_2},\bk_{b_3})\Big] +P^{t't'}(\bk_i)\,B_{2\delta}^{ttt'}(\bk_{b_1},\bk_{b_2},\bk_{b_3})\bigg\}\;, \nonumber \\
\mathrm{Cov}\l P_i^{tt'}, B_b^{t't't'}\r &= \l\delta_{ib_1}+\delta_{ib_2}+\delta_{ib_3}\r \,P^{tt'}(\bk_i)\,B^{t't't'}(\bk_{b_1},\bk_{b_2},\bk_{b_3})
\notag \\
&~~~+P^{t't'}(\bk_i)\,B_{1\delta}^{tt't'}(\bk_{b_1},\bk_{b_2},\bk_{b_3})\;, \nonumber \\
\mathrm{Cov}\l P_i^{t't'}, B_b^{ttt}\r &= 2P^{tt'}(\bk_i)\,B_{1\delta}^{ttt'}(\bk_{b_1},\bk_{b_2},\bk_{b_3})\;,  \nonumber \\
\mathrm{Cov}\l P_i^{t't'}, B_b^{(ttt')}\r &= \frac{2}{3}\,\Big[P^{t't'}(\bk_i)\,B_{1\delta}^{ttt'}(\bk_{b_1},\bk_{b_2},\bk_{b_3})+P^{tt'}(\bk_i)\,B_{2\delta}^{tt't'}(\bk_{b_1},\bk_{b_2},\bk_{b_3})\Big]\;, \nonumber \\
\mathrm{Cov}\l P_i^{t't'}, B_b^{(tt't')}\r &= \frac{2}{3}\,\Big[\l\delta_{ib_1}+\delta_{ib_2}+\delta_{ib_3}\r\, P^{tt'}(\bk_i)\,B^{t't't'}(\bk_{b_1},\bk_{b_2},\bk_{b_3}) \notag \\
&~~~+P^{t't'}(\bk_i)\,B_{2\delta}^{tt't'}(\bk_{b_1},\bk_{b_2},\bk_{b_3})\Big]\;, \nonumber \\
\mathrm{Cov}\l P_i^{t't'}, B_b^{t't't'}\r &= 2\,\l\delta_{ib_1}+\delta_{ib_2}+\delta_{ib_3}\r \,P^{t't'}(\bk_i)\,B^{t't't'}(\bk_{b_1},\bk_{b_2},\bk_{b_3})\;.
\end{align}
\vspace{3em}

Here the partially symmetric bispectra are given by:
\begin{align}
    B_{1\delta}^{ttt'}(\bk_{b_1},\bk_{b_2},\bk_{b_3})&=\delta_{ib_1}\,B^{t'tt}(\bk_{b_1},\bk_{b_2},\bk_{b_3})+\delta_{ib_2}\,B^{tt't}(\bk_{b_1},\bk_{b_2},\bk_{b_3}) \notag \\
    &~~~+\delta_{ib_3}\,B^{ttt'}(\bk_{b_1},\bk_{b_2},\bk_{b_3})\;,
\\
    B_{2\delta}^{ttt'}(\bk_{b_1},\bk_{b_2},\bk_{b_3})&=[\delta_{ib_1}+\delta_{ib_2}]\,B^{ttt'}(\bk_{b_1},\bk_{b_2},\bk_{b_3})+[\delta_{ib_1}+\delta_{ib_3}]\,B^{tt't}(\bk_{b_1},\bk_{b_2},\bk_{b_3})  \notag \\
&~~~+[\delta_{ib_2}+\delta_{ib_3}]\,B^{t'tt}(\bk_{b_1},\bk_{b_2},\bk_{b_3})\;.
\end{align}

These bispectrum tracer indices  are used for convenience and can be generalised for any 2-tracer combination, e.g.\ $tt't'$. Note that a nonzero power spectrum-bispectrum cross-covariance is generated by configurations in which the power spectrum $\bk$ mode coincides at least with 
one of the sides of a triangle configuration. The results presented above for the cross-covariance between power spectrum and bispectrum, consider only the dominant contribution, while the 5-point connected correlation function part is neglected (see \cite{Biagetti2022} for a discussion). Note that the block matrix of \cref{eq:blockPB} and each of the sub-blocks are not square matrices.

The expressions for 3 tracers are in \cref{app:3t}.

\section{Fisher information on $\fnl$}\label{sec:fisher}

The previous section presented the  formalism  for the multi-tracer power spectra, the multi-tracer bispectra and their combination. 
In this section, we use the formalism for a simplified qualitative analysis of the improvements from the multi-tracer  constraints on $\fnl$ relative to the single-tracer as baseline. In order to achieve this, we consider the same volume and redshift range for each single-tracer and their multi-tracer. This means that the improvement shown here on local PNG only refers to the effect of combining the samples or not, without taking into account the effect of different volumes and redshift overlap of two different samples. To this end,  we consider only the conditional errors, and we use nominal surveys without considering details and neglecting systematics. More realistic forecasts, that marginalise over relevant cosmological and nuisance parameters and use specifications for current and upcoming galaxy surveys, will be treated in a follow-up paper \cite{Karagiannis:2023}.
Here we also confine ourselves
to separate investigations into the precision  from the multi-tracer power spectra and from the multi-tracer bispectra -- without combining the information obtained from the two correlators. The follow-up paper \cite{Karagiannis:2023} will  include this combination of multi-tracers as part of more comprehensive forecasts.

The Fisher information on $\fnl$ {in each redshift bin}  for the multi-tracer power spectra  is given by 
  \begin{align} 
{F_{\fnl}^{\bm P}}&= \int {\rm d} \bk \,\frac{\partial\bm{D_P}(\bk)}{\partial \fnl}\,\bm{\mathrm{Cov}_P}^{-1}{(\bk)}\,\frac{\partial\bm{D_P}^{{\sf T}}(\bk)}{\partial\fnl}\;.
  \end{align}
For a rough estimate of the improvements on precision that are produced by the multi-tracer,
we consider the  error  
$\sigma^{\bm P}(\fnl)=(F^{\bm P}_{\fnl})^{-1/2}$. Then an analytic expression can be derived for  the single-tracer power spectrum: 
\begin{equation}\label{eq:analytic_PS_ST}
  \sigma^I_{\fnl}=N_P^{1/2}\,\int {\rm d}\bk \,  \frac{M(k)\,Z_1^I(\bk)}{2\,{b_\Psi^I}\, k^\alpha }\,\frac{{P^{I}}(\bk)}{P^{I}(\bk)-P_\veps^I}, \qquad N_P=2\,\frac{(2\,\pi)^3}{V_s\,V_k}\;,
  \end{equation}
where we omitted the redshift dependence.
Here  $\alpha=1,0$ for  orthogonal and local shapes, respectively \cite{Dalal2008,Slosar2008,Giannantonio2010,Schmidt2010,Giannantonio2012,Desjacques2016,Cabass:2018roz}. Note that equilateral PNG ($\alpha=2$) cannot be constrained using the galaxy power spectrum, as shown in \cite{Assassi2015}. Using the linear redshift space kernel \cref{eq:Z1}, the analytic result \cref{eq:analytic_PS_ST} shows that, for a fixed number density and $\mu$, the tightest constraints on $\fnl$ can be obtained from a sample that has $b_\Psi^I \gg b_1^I$.

For the multi-tracer case the analytic expression of the Fisher matrix on $\fnll$ is given by:
\begin{align}\label{eq:ffnl}
 F^{\bm P}_{\fnl}&= \sum_{I,J}\int {\rm d}\bk \, {\cal N}^{-1}(\bk)\,\bigg\{ {4\,P(k)^2(b_\Psi^J)^2\,\big[P_\veps^I+P_\veps^J\,R(\bk)^2\big]^2\,\left[Z_{1}^{J}(\bk)\right]^2}
 \nonumber \\
&+
{8\,P(k)^2\,P_\veps^J\,b_\Psi^J\,\left[Z_{1,\rm G}^J(\bk)\,b_\Psi^I-Z_{1,\rm G}^I(\bk)\,b_\Psi^J\,\right]\,\big[P_\veps^I+P_\veps^J\,R(\bk)^2\big]\,R(\bk)\,Z_{1}^{J}(\bk)} 
\nonumber \\
&+
 {2\, P(k)^2\left[Z_{1,\rm G}^J({\bk})\,b_\Psi^I-Z_{1,\rm G}^I({\bk})\,b_\Psi^J\right]^2\,\big[P_\veps^I\,P^{J}(\bk)+P_\veps^J\,\big\{ P_\veps^J+P^{J}(\bk)\big\}\,R^2{(\bk)}\big]} \bigg\}\;,%
\end{align}
where
\begin{align}
{\cal N}(\bk)= N_P\,M(k)^2
\,k^{-2\alpha}\,\big[P_\veps^I\, P^{J}(\bk)+P_\veps^J \,\big\{ P^{J}(\bk)-P_\veps^J \big\}\, R(\bk)^2\big]^2\quad\mbox{with}~
R(\bk)=\frac{Z_{1}^{I}(\bk)}{Z_{1}^{J}(\bk)}\;.
\end{align}
Here $Z_{1,\rm G}^I(\bk)$ is the Gaussian version of \cref{eq:Z1}, i.e.\ for $\fnl=0$. Note that this expression is valid for a general linear redshift kernel (e.g.\ including non-integrated relativistic corrections). 

Our \cref{eq:ffnl} generalises the result of \cite{Barreira:2023rxn}, which only considers local PNG, takes the cosmic variance limit and neglects RSD effects. At fixed redshift bin and scales, \cref{eq:ffnl} indicates that precision on $\fnl$ is increased by
\begin{equation}\label{eq:ffnl2}
\mbox{increasing}~~\big|Z_{1,\rm G}^J({\bk})\,b_\Psi^I-Z_{1,\rm G}^I({\bk})\,b_\Psi^J\big|~~ \mbox{and} ~~ R(\bk).
\end{equation}
One way to maximise the former is by choosing appropriately the line-of-sight angle $\mu$ to gain a boost from the presence of RSD, which is also evident in forecasts that utilise single tracers (e.g.\ \cite{Karagiannis2018}). If we neglect RSD (or fix  $\mu$) and consider local PNG, then \cref{eq:ffnl2} implies maximising $|b_1^J\,b_\Phi^I-b_1^I\,b_\Phi^J|$, as shown also in \cite{Barreira:2023rxn}. 

For the multi-tracer bispectra, the Fisher matrix is
\begin{align}
  {F_{\fnl}^{\bm B}}&= 
  {\int {\rm d}^3\bk}
  \frac{\partial\bm{D_B}(\bk_1,\bk_2,\bk_3)}{\partial\fnl}\,\bm{\mathrm{Cov}_B}^{-1}
{(\bk_1,\bk_2,\bk_3)}\,\frac{\partial\bm{D_B}^{{\sf T}}(\bk_1,\bk_2,\bk_3)}{\partial\fnl}\;.  
\end{align}
Analytical calculations for the bispectrum multi-tracer are long and uninformative, without assuming some limit, either for the shot noise or the triangle configurations. 
Consequently, we  perform the bispectrum analysis numerically. 

We consider two nominal spectroscopic surveys, with the following features:\\ $\bullet$ Overlap sky area = $10\,000\,\deg^2$, overlap redshift range  $0.9\leq z\leq 1.7$ (bin size $\Delta z=0.1$).\\
$\bullet$
\underline{Tracer $t$}:

high density,  $\order{10^{-3}}\;(\Mpc)^3$; 
$\bar n_t(z)$
similar to a Euclid sample \cite{Euclid:2019clj}; 

linear bias $b_1^t(z)=1.3+0.6\,z$. \\
$\bullet$
\underline{Tracer $t'$}:

lower density, $\order{10^{-4}}\;(\Mpc)^3$; 
$\bar n_{t'}(z)$ similar to DESI sample \cite{DESI:2016fyo};  

linear bias $b_1^{t'}(z)=0.84/D(z)$. \\
In order to derive the higher-order and non-Gaussian bias coefficients needed for the tree-level bispectrum, we use a simplified halo-occupation distribution (HOD) model from \cite{Yankelevich:2018uaz}. The HOD model has two free parameters which are calibrated from the values of the linear bias and number densities at each redshift. 

This set-up for the nominal surveys exploits the findings of \cref{eq:ffnl} and will function as the benchmark case. The fiducial cosmology is a flat $\Lambda$CDM model, with $\omega_c=0.1199$, $\omega_b= 0.02237$, $h=0.6736$, $n_s=0.96488$, $A_s=2.1\times10^{-9}$, as measured by \cite{Planck2018_cosmo}. In addition, we assume non-Gaussian initial conditions, with an amplitude of $\fnl=20$, for all three PNG types considered here. The scales used to perform the Fisher matrix analysis are well within the perturbative regime: $k_\text{min}(z)= k_{\rm f}(z)\equiv2\,\pi/V_s(z)^{1/3}$ and $k_\text{max}(z)=0.75\,k_\text{NL}(z)$, 
{where $k_{\rm f}(z)$ and $V_s(z)$ are the fundamental frequency and the volume of the survey within a given redshift bin, respectively}, and $k_\text{NL}$ is given by the inverse square root of the one-dimensional velocity dispersion (see \cite{Karagiannis:2019jjx}; a plot of $k_\text{NL}(z)$ is shown in \cite{Karagiannis:2022ylq}).

\subsection*{\underline{$\sigma(\fnl)(z)$: benchmark case}}
\begin{figure}[t]
\centering
\hspace*{-1.3cm}
\resizebox{1.1\textwidth}{!}{\includegraphics{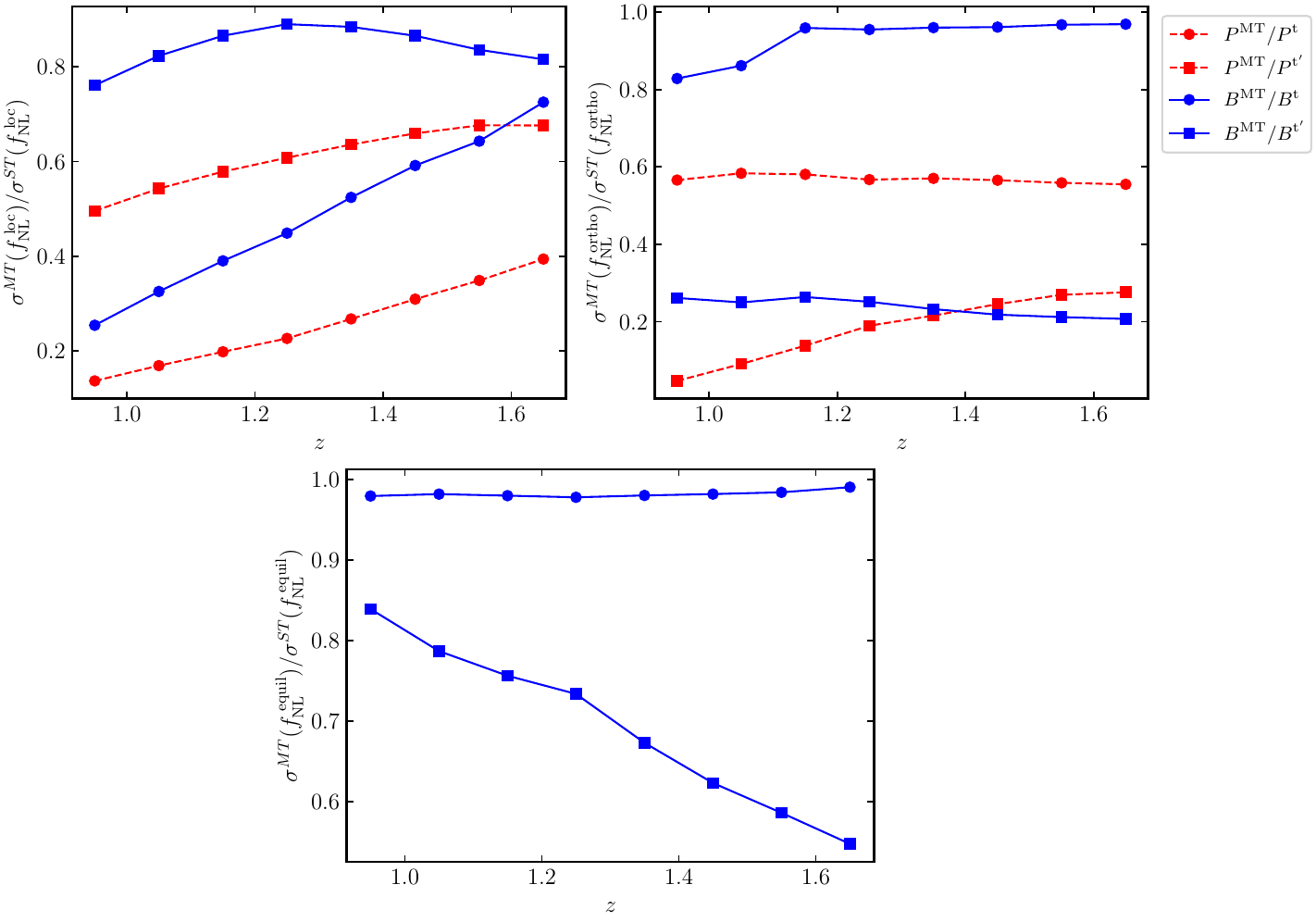}}  
\caption{
Relative change in the cumulative $1\sigma$ error on $\fnl$, provided by the multi-tracer (MT) approach relative to each single tracer (ST), as a function of redshift. Power spectrum (red lines) and bispectrum (blue lines) results, are shown for three PNG types. $t$ and $t'$ are tracers of the two nominal next-generation surveys considered here.
}
\label{fig:cumsigfnl_Euclid_DESI_ELG}
\end{figure}

\vspace*{0.1cm}
\cref{fig:cumsigfnl_Euclid_DESI_ELG}  presents the relative change in the conditional $\fnl$ errors from the multi-tracer approach with respect to the two nominal single-tracers $t$ and $t'$, as a function of redshift. The results are for three types of PNG, i.e.\ local, equilateral and orthogonal, using the power spectrum and bispectrum. For both correlators, the multi-tracer delivers a noticeable overall improvement relative to each single-tracer survey. 

In the case of local PNG, the multi-tracer power spectrum improvement over the $\fnl$ errors from a single tracer is significant -- well above $30\%$ across the redshift range. For the bispectrum, this applies only in the low-redshift bins of tracer $t$, while for tracer $t'$ the gain is more modest. However, the overall enhancement of the bispectrum multi-tracer approach is still important and in the range  $20-30\%$.

For orthogonal PNG, the improvement from the multi-tracer with respect to tracer $t'$ is equivalent for both correlators ($\sim70\%$), while regarding tracer $t$ it is significantly more for $P^{\rm MT}$ than for $B^{\rm MT}$. Note that the performances of the single tracers $t$ and $t'$ are opposite to the local case. \cref{eq:analytic_PS_ST} shows that the single-tracer conditional error from the power spectrum on $\fnl$ is parametrically determined by  $b_1,b_\Psi$ and $P_\varepsilon$. Numerical tests and the analytic results show that the constraints are more sensitive to the values of $b_\Psi$, in particular for the case where $b_\Psi^I \gg b_1^I$. For the local case, $b_\Psi^{t'} > b_\Psi^{t}$ and hence we see that tracer $t$ outperforms the constraints from $t'$, for both correlators. On the other hand, for orthogonal PNG, due to the nature of this shape (see \cite{Desjacques2016} for a discussion), $b_\Psi^{t} > b_\Psi^{t'}$, leading to an opposite behaviour with respect to the local PNG.

For equilateral PNG, the multi-tracer bispectrum gives only a marginal improvement over the performance of tracer $t$, as in the orthogonal case. On the other hand, the gain for tracer $t'$ increases with redshift, reaching a substantial $\sim50\%$. Note that the power spectrum does not provide any constraints on $\fnle$ (see e.g.\ \cite{Assassi2015}).

The attributes of the nominal surveys were selected to produce a large  $|b_1^J\,b_\Psi^I-b_1^I\,b_\Psi^J|$.
The results in \cref{fig:cumsigfnl_Euclid_DESI_ELG} indicate that in the general benchmark case -- namely, with no additional assumptions -- this is enough to achieve a significant gain from the multi-tracer approach in most cases, but for the rest it is insufficient.

Apart from the dependence on  PNG type shown in \cref{fig:cumsigfnl_Euclid_DESI_ELG}, additional parameters have a significant effect on the level of improvement  achieved by the multi-tracer. The main tracer-dependent variables that affect $\sigma(\fnl)$, besides the linear and PNG biases ($b_1^I$ and $b_\Psi^I$), are the number densities $\bar{n}_I$, which in turn affect the stochastic contributions in \cref{stoch-st}. In addition, for the bispectrum, higher-order bias parameters such as $b_2^I$ will have an effect on the constraining power.

Further examination is needed into the parametric behaviour of the multi-tracer power spectrum and bispectrum, in order to determine the tracer properties that maximise gain from combining their information.  Here we  confine attention to  local PNG, and we fix the redshift at $z=1$. Note that $b_\Psi\rightarrow b_\Phi$ for local PNG, as follows from \cref{psiphi} with $\alpha=0$.

 \begin{figure}[t]
\centering
\resizebox{\textwidth}{!}{\includegraphics{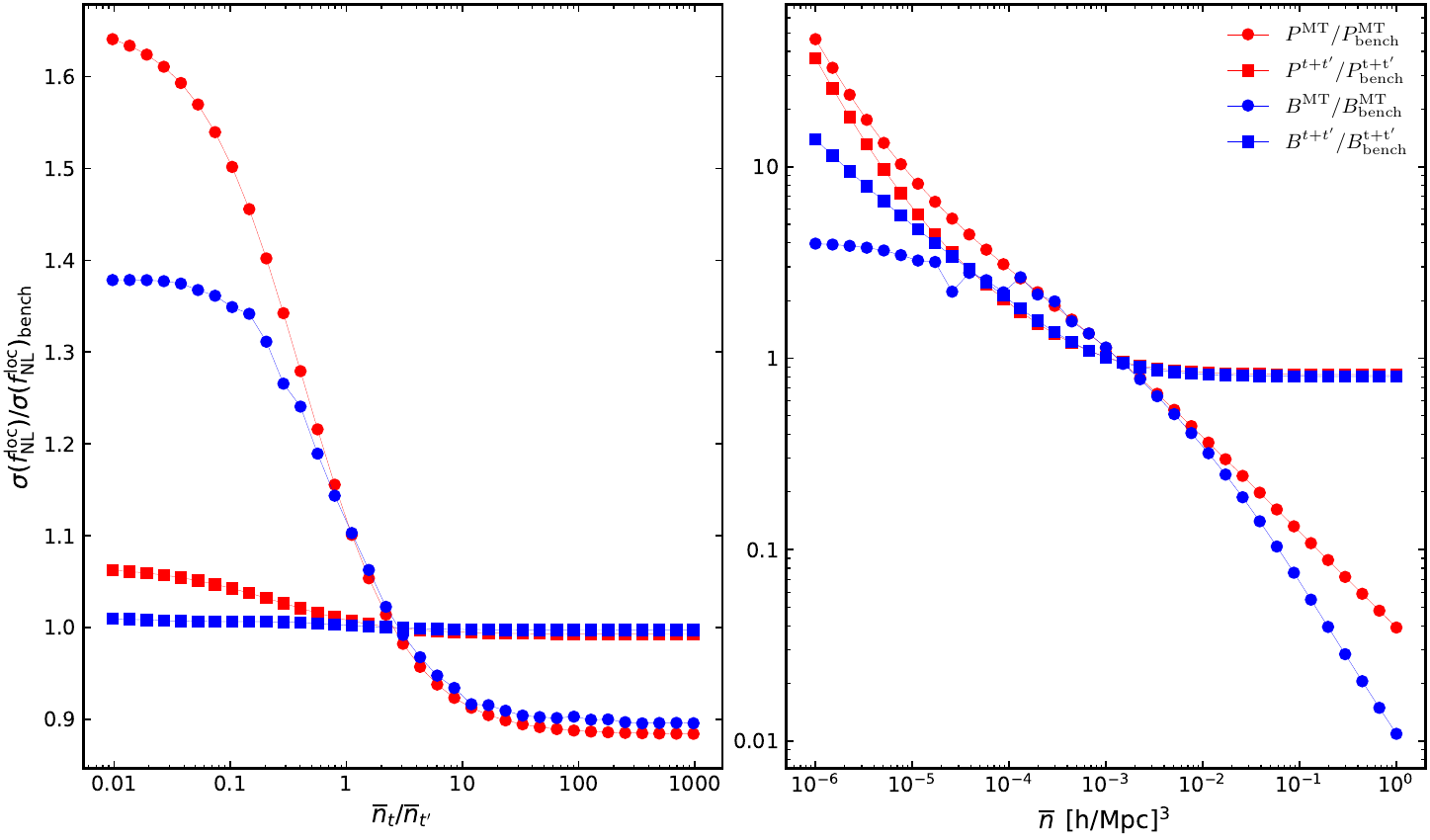}}  
\caption{
{\it Left:}  The dependence of the $1\sigma$ conditional $\fnll$ errors on the ratio $\bar{n}_t/\bar{n}_{t'}$, at fixed redshift $z=1$. We vary only the number density of sample $t$, while fixing all other parameters of tracer $t$, together with those of sample $t'$. {\it Right:} Testing the cosmic variance limit, by enforcing equal number densities, i.e.\ $\bar{n}_t=\bar{n}_{t'}=\bar{n}$, and varying $\bar n$ (at $z=1$). Both panels show the multi-tracer forecasts (circles) and the results from the simple combination of the two Fisher matrices (squares) (without considering the cross-terms in the data vectors and block covariances), divided by the benchmark results of \cref{fig:cumsigfnl_Euclid_DESI_ELG} at $z=1$, for both summary statistics. }
\label{fig:sigfnl_N1oN2_N}
\end{figure} 

\subsection*{\underline{$\sigma(\fnll)$: effect of number densities}}
\vspace*{0.1cm}
We start with the effect of number densities on  $\fnll$ conditional errors. We vary $\bar{n}_t$, while fixing all other parameters of tracers $t$ and $t'$, including $\bar{n}_{t'}$. The dependence of the $\fnll$ errors on the ratio $\bar{n}_t/\bar{n}_{t'}$, at $z=1$, is shown in the left panel of \cref{fig:sigfnl_N1oN2_N}. In addition to the multi-tracer results (marked with the superscript MT), we present the forecasts from simply adding the Fisher matrices of the two tracers (marked with the superscript $t+t'$), without considering the cross-terms in the data vectors, \cref{eq:PSdata,,eq:BSdata}, and block covariances, \cref{eq:PS_cov,,eq:BS_cov}. 

The left panel of \cref{fig:sigfnl_N1oN2_N} shows that for both correlators the multi-tracer gives maximal improvement on $\fnll$ errors for $\bar{n}_t/\bar{n}_{t'}
\gtrsim 10$. In other words, the number density of one tracer should be
at least an order of magnitude larger than that of the other. In fact, if the ratio is beyond that point, the constraints do not improve further, as we remain shot-noise dominated by the low-density tracer. This behaviour occurs since only one of the two tracers is allowed to reach the limit where stochastic contributions can be neglected (i.e.\ the cosmic variance limit). The other tracer has a constant shot-noise component, that ultimately saturates the multi-tracer Fisher information. Nonetheless, in this setup, the multi-tracer power spectrum seems to be more sensitive to the cosmic variance limit of tracer $t$, than the bispectrum. This is an expected behaviour, since the effect of local PNG appears only on the largest scales for the tracer power spectrum, whereas the effect on the tracer bispectrum is from the squeezed  configurations that are not purely dependent on the largest scales, since they couple large scales to small scales.

On the other hand, the constraints from just summing the Fisher matrices of the two tracers (i.e.\ neglecting the multi-tracer cross-terms), exhibit a minimal change as a function of the number density ratio. This indicates that the complete multi-tracer approach takes advantage of the increasing number density in an optimal way, by maximising the contribution of the cross-terms and hence providing a significant improvement on the $\fnll$ constraints, i.e.\ up to $10\%$ for $\bar{n}_t/\bar{n}_{t'}
\gtrsim 10$. Moreover, for the low density regime ($\bar{n}_t/\bar{n}_{t'}
< 1$), neglecting the multi-tracer cross-terms, when considering the summed information from two overlapping tracers, can lead to severe overestimation of the $\fnll$ errors (even up to $40-60\%$), in the case of both correlators.

 Note that the scenario with equal number densities of tracers  (i.e.\ $\bar{n}_t/\bar{n}_{t'}=1$) is not an optimal set-up for a multi-tracer application, at least when neither of the tracers approaches the high-density limit.

In order to investigate further the limit where the shot noise can be neglected, we assume equal number densities. Our aim is to explore the low- and high-density limits. The dependence of the forecast error on $\fnll$  as a function of number density is presented in the right panel of \cref{fig:sigfnl_N1oN2_N}, for both correlators. This indicates that the multi-tracer constraints on local PNG exhibit an inverse power-law behaviour as the number densities of the tracers approach the cosmic variance limit,  $\bar n\gg 10^{-3}\;(\Mpc)^3$. By contrast, the results of simply adding the Fisher matrices of the two tracers saturate around the same value, without exhibiting any further improvement as the number densities increase. This shows the clear advantage of the multi-tracer approach, where the cross-term contribution is enhanced substantially, due to the decreasing shot-noise terms, providing significant improvement on $\fnll$.  

Our power spectrum results clearly show the benefit of the multi-tracer, for the high density samples, as well as for tracers that approach the cosmic variance limit. This agrees with the findings of the original work on the multi-tracer approach \cite{Seljak:2008xr}, as well as the more recent results of \cite{Barreira:2023rxn}. Moreover, here we show that the advantage of the multi-tracer towards the high-density limit can also be extended to the bispectrum, which exhibits a significant improvement compared to the power spectrum in the cosmic variance limited regime.

 \begin{figure}[t]
\centering
\resizebox{0.6\textwidth}{!}{\includegraphics{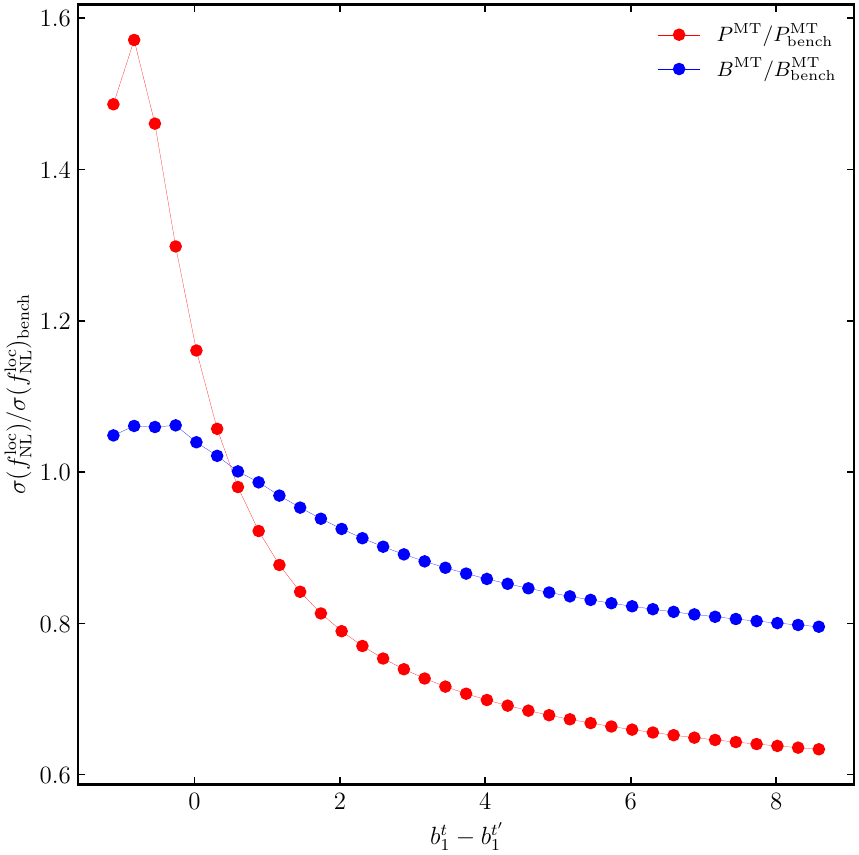}}  
\caption{
The dependence of the $1\sigma$ conditional $\fnll$ error on the linear bias difference $b_1^t-b_1^{t'}$, at  fixed redshift $z=1$, for the multi-tracer power spectrum and bispectrum. Here we vary only $b_1^t$, while fixing all other tracer-dependent parameters for both $t$ and $t'$. The results are normalised to the benchmark forecasts of \cref{fig:cumsigfnl_Euclid_DESI_ELG}.}
\label{fig:sigfnl_b1Imb1J}
\end{figure}

\subsection*{\underline{$\sigma(\fnll)$: effect of linear biases}}
\vspace*{0.1cm}
Next we investigate the effect of the linear biases on the multi-tracer power spectrum and bispectrum.  Due to the symmetrisation of the two correlators under tracer permutations, the outcome of the parametric analysis does not depend on which tracer's parameters we chose to vary. 
Note that the higher-order bias coefficients of each tracer depend on the HOD model, whose free parameters are calibrated by the value of the linear bias at each redshift. This means that varying $b_1^t$ will, in the HOD framework, affect all the bias coefficients of tracer $t$. Here we are interested in isolating the effect of a change in linear bias and we vary only  $b^1_t$ for tracer $t$, while fixing all other parameters for  tracers $t$ and $t'$. 

The relative change of the multi-tracer constraints on $\fnll$ over the benchmark case, for the two correlators and at $z=1$, are presented in \cref{fig:sigfnl_b1Imb1J} as a function of the difference between the linear biases, $b_1^t-b_1^{t'}$.  The results indicate that when the difference between the two biases is $\gtrsim 2$, there is a significant improvement (up to $\sim30\%$) with respect to the benchmark results, $\sigma(\fnll)/\sigma(\fnll)_{\rm bench}=1$, for the multi-tracer power spectrum. For the multi-tracer bispectrum, the gain is more modest. Beyond that point, the improvement slowly saturates ($|b_1^t-b_1^{t'}|>5$) for both correlators. The results show that the constraining power on $\fnll$ is $\propto b_1^t-b_1^{t'}$. Note that the case where the linear biases of the tracers are similar, $|b_1^t- b_1^{t'}|\sim0$, corresponds to the least tight constraints on $\fnll$, for both correlators. In this case, the power spectrum constraints degrade significantly ($\sim60\%$) compared to the bispectrum ($\sim 10\%$).

\subsection*{\underline{$\sigma(\fnll)$: effect of $b_1^{t'}\,b_\Phi^t-b_1^t\,b_\Phi^{t'}$}}
\vspace*{0.1cm}

\begin{figure}[t]
\centering
\resizebox{0.6\textwidth}{!}{\includegraphics{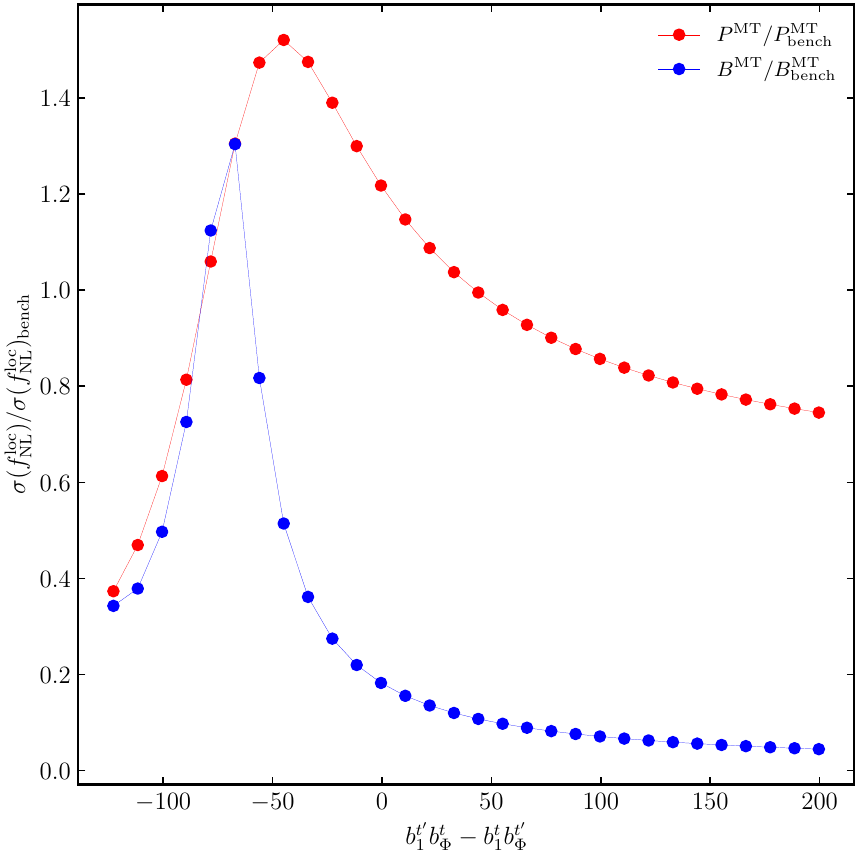}}  
\caption{
Same as \cref{fig:sigfnl_b1Imb1J}, where we not only vary $b_1^t$, but we allow all the other bias parameters of tracer $t$ to change in accordance with the dependence of the HOD model on the linear bias, including $b_\Phi^t$. The relative change, with respect to the benchmark results, is shown as a function of $b_1^{t'}\,b_\Phi^t-b_1^t\,b_\Phi^{t'}$. }
\label{fig:sigfnl_b1IbphiJmb1JbphiI}
\end{figure}

We proceed by allowing all other bias coefficients of tracer $t $ to vary as well, through the dependence of the HOD free parameters on the $b_1^t$ value, including $b_\Phi^t(b_1^t)$ via the universality relation assumption. The relative change of the $\fnll$ forecasts, compared to the benchmark results, as a function of $b_1^{t'}\,b_\Phi^t-b_1^t\,b_\Phi^{t'}$, is displayed in \cref{fig:sigfnl_b1IbphiJmb1JbphiI}. The behaviour exhibited indicates that for both correlators, the gain on the $\fnll$ forecasts increases as a function of $|b_1^{t'}\,b_\Phi^t-b_1^t\,b_\Phi^{t'}|$, reaching significant levels.

For the multi-tracer power spectrum, the improvement on $\fnll$ constraints -- i.e., the case where \(\sigma(\fnll)/\sigma(\fnll)_{\rm bench}<1\) -- saturates at $\sim20\%$ when dealing with large positive values of $b_1^{t'},b_\Phi^{t}-b_1^{t},b_\Phi^{t'}$. The maximum gain seems to be achieved towards the negative values of $b_1^{t'},b_\Phi^{t}-b_1^{t},b_\Phi^{t'}$. The improvement towards the negative values is in fact significantly larger than the saturation regime towards the large positive values. Nonetheless, maximising $|b_1^{t'},b_\Phi^{t}-b_1^{t},b_\Phi^{t'}|$ provides the highest gain from the multi-tracer power spectrum, verifying the analytical findings of \cref{eq:ffnl}.

For the multi-tracer bispectrum, the tightest constraint on $\fnll$ is provided once $b_1^{t'}\,b_\Phi^t-b_1^t\,b_\Phi^{t'}>0$. Beyond that, the improvement relative to the benchmark saturates. This indicates the different behaviour of the two correlators as a function of $b_1^{t'}\,b_\Phi^t-b_1^t\,b_\Phi^{t'}$, once we allow for all bias parameters of tracer $t$ to change and $b_\Phi^t\propto b_1^t$.  Moreover, the multi-tracer power spectrum results for  $b_1^{t'}\,b_\Phi^t-b_1^t\,b_\Phi^{t'}\approx0$ exhibit a degradation ($\sim20\%$) relative to the benchmark case, while for the bispectrum the constraints are close to the optimal outcome.

\begin{figure}[t]
\centering
\resizebox{\textwidth}{!}{\includegraphics{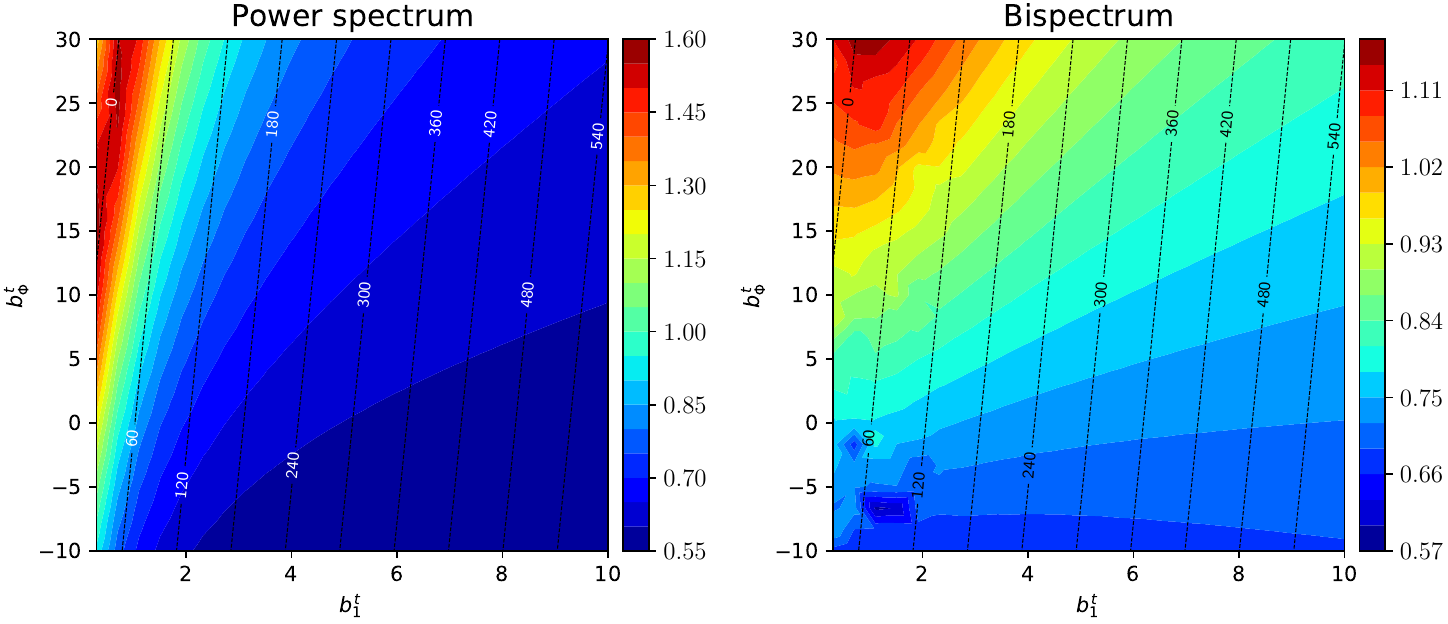}}  
\caption{The ratio $\sigma(\fnll)/\sigma(\fnll)_{\rm bench}$, with values given by the colour bar, in the $(b_\Phi^t,\,b_1^t)$ plane, for multi-tracer power spectrum (\emph{left}) and bispectrum (\emph{right}). We vary $b_1^t$ and $b_\Phi^t$ independently, without assuming a universality relation, while fixing all other parameters for both tracers.  Black dotted lines are contours of constant  $b_1^{t'}\,b_\Phi^t-b_1^t\,b_\Phi^{t'}$. }
\label{fig:contour}
\end{figure}

The different parametric behaviour of the multi-tracer power spectrum and bispectrum can also be seen in the contour plots of \cref{fig:contour}. In this case, we vary $b_1^t$ and $b_\Phi^t$, without assuming any relation between the two (i.e.\ no universality relation), while fixing all other tracer-dependent parameters for both tracers. The black dotted contours are lines of constant $b_1^{t'}\,b_\Phi^t-b_1^t\,b_\Phi^{t'}$. The results show that  for the multi-tracer power spectrum, the case $b_1^{t'}\,b_\Phi^t-b_1^t\,b_\Phi^{t'}\approx 0$ corresponds to the weakest constraints, where errors increase compared to the benchmark case. The maximal improvement on the forecast precision originates from the case of maximal $b_1^{t'}\,b_\Phi^t-b_1^t\,b_\Phi^{t'}$ values -- in particular for  $-10\le b_\Phi\le10$ and for $b_1\ge 2$. The behavior of the multi-tracer bispectrum is somewhat similar, but in the $b_1^{t'}\,b_\Phi^t-b_1^t\,b_\Phi^{t'}\approx 0$ regime, which corresponds also to weaker constraints than the benchmark, the performance is better than the power spectrum equivalent. As shown in \cref{fig:sigfnl_b1IbphiJmb1JbphiI}, the optimal constraints from the multi-tracer bispectrum arise for an increasing $|b_1^{t'}\,b_\Phi^t-b_1^t\,b_\Phi^{t'}|$. In particular, in order to achieve a significant improvement, growing  $b_1^t$ values and  very negative $b_\Phi^t$ are required (blue shaded region in the right panel of \cref{fig:contour}).

The results presented in \cref{fig:sigfnl_b1Imb1J,,fig:sigfnl_b1IbphiJmb1JbphiI,fig:contour} indicate the potential of the multi-tracer to significantly improve  constraints on local PNG by carefully choosing the two tracers. For the multi-tracer power spectrum and bispectrum, increasing $|b_1^t-b_1^{t'}|$ or $|b_1^{t'}\,b_\Phi^t-b_1^t\,b_\Phi^{t'}|$ gives the optimal results, where the latter leads to overall substantial enhancement.

\subsection*{\underline{$\sigma(\fnll)$: effect of second-order biases}}
\begin{figure}[t]
\centering
\resizebox{0.65\textwidth}{!}{\includegraphics{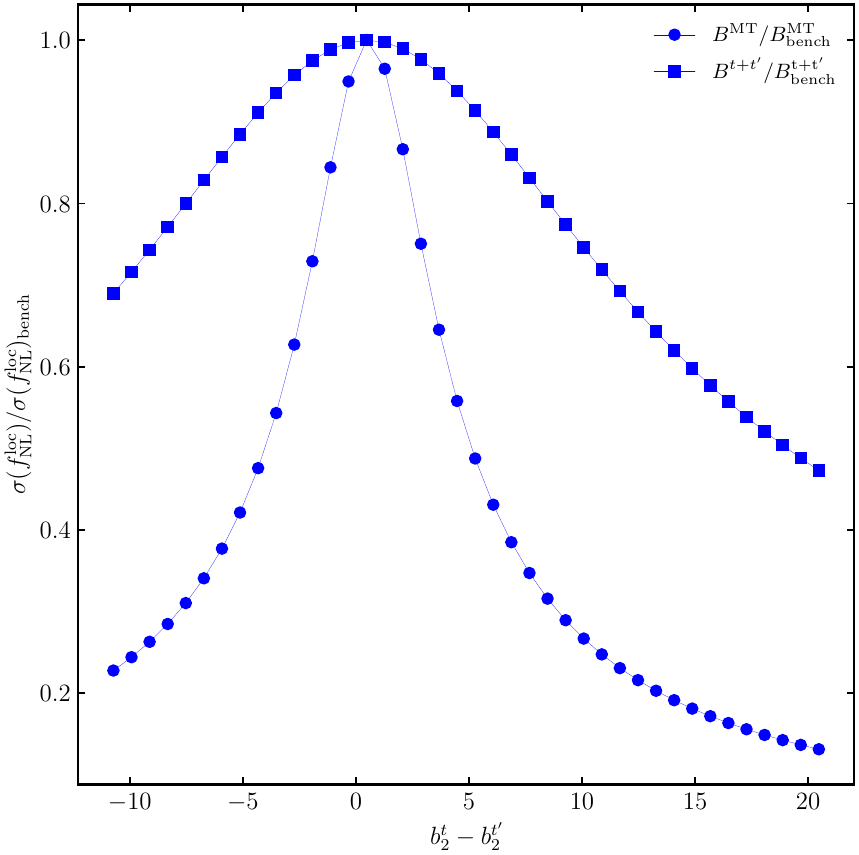}}  
\caption{
$\sigma(\fnll)/\sigma(\fnll)_{\rm bench}$ as a function of the difference $b_2^t-b_2^{t'}$, varying $b_2^t$, while keeping fixed all other parameters for both tracers. Square points correspond to the summation of the two Fishers,  while  dots represent the full multi-tracer results. 
}
\label{fig:sigfnl_b2}
\end{figure}

\vspace*{0.1cm}
In order to investigate further the parametric behaviour of the multi-tracer tree-level bispectrum, we vary the second-order bias coefficient $b_2^t$, while all other parameters for both tracers are fixed. The relative improvement results are shown in \cref{fig:sigfnl_b2} as a function of the difference $b_2^t-b_2^{t'}$ for the complete multi-tracer approach, and for the sum of the individual Fisher matrices. Comparing  the two shows once again the importance of utilising the full multi-tracer  in order to optimise the gain on $\fnll$ constraints compared to the summed information of the bispectrum of two tracers.

In addition, varying $b_2^t$ shows a peak around the case where the nonlinear bias parameters of tracers $t$ and $t'$ are similar, i.e.\ the difference $|b_2^t-b_2^{t'}|$ is small. This indicates the regime where the multi-tracer bispectrum offers minimal improvement with respect to the benchmark results. Therefore, the constraining power on $\fnll$ is $ \propto|b_2^t-b_2^{t'}|$. In fact, any value outside of the range $-1<|b_2^t-b_2^{t'}|<5$ can have an improvement of $>20\%$. Moreover, increasing the value of $|b_2^t-b_2^{t'}|$ improves the gain on $\fnll$ as much as, or even more than, the case of an increasing $|b_1^t-b_1^{t'}|$ or $|b_1^{t'}\,b_\Phi^t-b_1^t\,b_\Phi^{t'}|$ (see \cref{fig:sigfnl_b1Imb1J,,fig:sigfnl_b1IbphiJmb1JbphiI}). This reveals the importance of the nonlinear bias parameter in utilising the full potential of the multi-tracer bispectrum.

\subsection*{\underline{$\sigma(\fnll)$: effect on scales}}

The benefits of the multi-tracer approach for local PNG constraints beyond the nominal scenario studied here, should be explored. Although this is left for a future work, we can provide in this section some preliminary indication of whether the improvement shown in \cref{fig:cumsigfnl_Euclid_DESI_ELG} would persist in any upcoming survey. To achieve this we investigate the sensitivity of the results to the minimum and maximum accessible scales. This is done in order to explore whether or not the multi-tracer gain originates from the expected scale range -- and hence if it is likely to persist in a realistic set-up. 

The information on the amplitude of local PNG, in the case of the power spectrum, comes from the scale-dependent correction to the linear bias (see \cref{sec:stpb}), while for the bispectrum it mostly originates from  squeezed configurations that couple long and short modes. The latter has an additional source of information, similar to the scale-dependent bias correction in the power spectrum (see e.g.\ \cite{Karagiannis2018}). 
While the power spectrum PNG constraints are very sensitive to these scales, the bispectrum can still provide competitive results in the absence of the very large scales, as long as enough squeezed triangles are probed by the survey \cite{Karagiannis2018,Karagiannis:2019jjx,Karagiannis:2020dpq}.

\begin{figure}
    \centering
    \includegraphics[width=\textwidth]{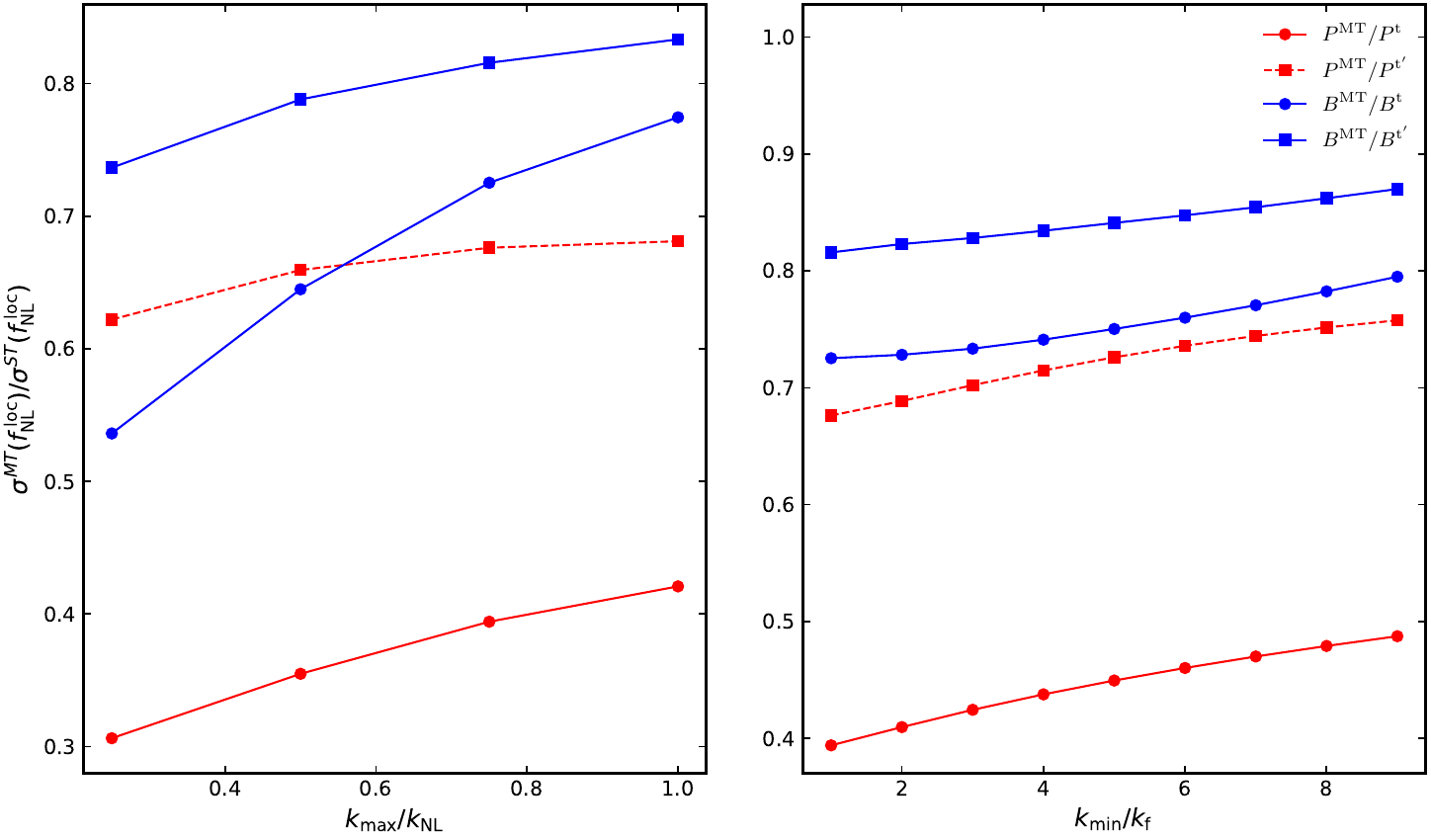}
    \caption{The relative change of the multi-tracer compared to single tracers $t$ and $t'$, for the power spectrum (red) and bispectrum (blue), as a function of the maximum wavenumber $k_{\rm max}$ (left panel) and minimum $k_{\rm min}$ (right panel). The $\fnll$ errors correspond to the cumulative contribution from all redshift bins, while $k_{\rm NL}$ and $k_{\rm f}$ are given in the text under \cref{fig:cumsigfnl_Euclid_DESI_ELG}. Note that the benchmark results in \cref{fig:cumsigfnl_Euclid_DESI_ELG} correspond to $k_{\rm max}/k_{\rm NL}=0.75$ and $k_{\rm min}/k_{\rm f}=1$ respectively.}
    \label{fig:sigfnl_kmin_kmax}
\end{figure}

The sensitivity of the multi-tracer improvement in  $\fnll$ constraints, as a function of the largest accessible scales $k_{\rm min}(z)$, is presented in the right panel of \cref{fig:sigfnl_kmin_kmax}. Here the expression for $k_{\rm max}(z)$ remains the same as in the benchmark case. The results are presented for both correlators, where the 1$\sigma$ errors originate from the cumulative information over the whole redshift range and each point assumes a fixed value of the ratio $k_{\rm min}(z)/k_{\rm f}(z)$.

In the case of the power spectrum, the relative gain of the multi-tracer with respect to the single-tracer exhibits only a small loss (up to $\sim10\%$ reduction), once we reduce the access to the large scales (large values of the ratio $k_{\rm min}(z)/k_{\rm f}(z)$). Although power spectrum constrains on $\fnll$ are the most sensitive to these scales, the multi-tracer approach successfully maintains the relative improvement over single-tracer results, at more or less the same level. 

The multi-tracer bispectrum follows a similar trend, where a $\sim 10\%$ reduction to the benchmark ($k_{\rm min}(z)/k_{\rm f}(z)=1$) improvement is observed. This decrease can be important for surveys like those considered here, where a maximum $20-30\%$ gain from the multi-tracer bispectrum over the single-tracer (see \cref{fig:cumsigfnl_Euclid_DESI_ELG}) is provided. However, in the case where the parametric behaviour (discussed in the previous subsections)  is used to advantage in the tracer selection process, the multi-tracer bispectrum could still provide a significant improvement to the local PNG constraints.

To investigate further the sensitivity of the multi-tracer improvement seen in the benchmark case to the small clustering scales, we vary the  ratio $k_{\rm max}(z)/k_{\rm NL}(z)$, while fixing $k_{\rm min}(z)$ to the fiducial choice. The results are presented in the left panel of \cref{fig:sigfnl_kmin_kmax}. The power spectrum exhibits a reduction, in the overall gain, that does not surpass $\sim10\%$. The constraints on $\fnll$ in this case do not depend on the information within the small scales. The observed decrease in the relative gain is simply attributed to the improvement of the $\fnll$ constraints from each single-tracer, while the multi-tracer errors are more or less unaffected. 

On the other hand, limiting substantially the accessibility to  small scales, i.e. in the regime $k_{\rm max}(z)/k_{\rm NL}(z)<0.5$, reduces the number of squeezed triangles, affecting the performance of the bispectrum in constraining $\fnll$. This is true for both the multi-tracer and single-tracer analysis. However, in this case we see that the gain of the multi-tracer bispectrum is greater, managing to maintain the relative improvement relative to each single tracer at significant levels (within the margin  $20$--$50\%$). This means that reducing the small-scale access affects the single-tracer analysis substantially more than the multi-tracer. In other words, the multi-tracer bispectrum does much better in a regime which is not shot-noise dominated. The gain exhibited reaches the benchmark values as soon as the small scale access is restored to the fiducial choice (i.e.\ $k_{\rm max}(z)/k_{\rm NL}(z)\gtrsim0.75$).

The overall trends suggests that the $\fnll$ constraints behave differently in the single and multi-tracer case. The single-tracer case is more sensitive to $k_{\rm max}$ than the multi-tracer, which leads to a larger improvement at lower $k_{\rm max}$ cuts. This indicates the significance of the multi-tracer approach, and especially for the bispectrum, in upcoming surveys. In cases where the small scales are lost due to the survey's design and systematics, 
the multi-tracer approach becomes ever more important.

\section{Summary and conclusion}

In this work we investigate the potential of the multi-tracer power spectrum and  multi-tracer  bispectrum analysis in improving the constraints on the amplitude of PNG, with a particular focus on local PNG. We consider two nominal next-generation surveys and use the Fisher matrix to probe the information on $\fnll$, within the power spectra and bispectra of the two tracers. The main goal is to quantify the impact of the various tracer-dependent parameters on the constraining power on $\fnll$, in order to maximise the gain from the multi-tracer technique and provide the tracer-selection criteria for future galaxy surveys. This is done by fixing the survey specifications for each single-tracer and the multi-tracer, which effectively means comparing the potential gain from combining the samples or not over the same volume and redshift range. The main findings can be summarised as follows:

\begin{itemize}
    \item The complete multi-tracer approach takes advantage of  increasing number density in an optimal way, by maximising the contribution of the cross-terms. This is particularly important for high-density samples, as well as for tracers that approach the cosmic variance limit. The scenario where the number densities of the tracers are similar does not provide the tightest constraints, except for high-density samples.
    
    \item The constraining power of the power spectrum and bispectrum multi-tracers is $\propto |b_1^{t}\,b_\Phi^{t'}-b_1^{t'}\,b_\Phi^{t}|$, where the power spectrum exhibits a much larger gain compared to the bispectrum. In particular the optimal regime is for an increasing difference between the linear biases of the two tracers, as well as for a negative $b_\Phi$ for one of the two tracers.

    \item Maximising the quantity $|b_1^{t'}\,b_\Phi^t-b_1^t\,b_\Phi^{t'}|$ is not enough for to fully exploit the potential of the multi-tracer bispectrum. Increasing as well the second-order bias difference $|b_2^t-b_2^{t'}|$  provides a significant gain in the bispectrum constraints, compared to only increasing $|b_1^{t'}\,b_\Phi^t-b_1^t\,b_\Phi^{t'}|$.

    \item The findings of this work demonstrate a substantial improvement in constraining $\fnl$ through the utilization of the multi-tracer bispectrum. However, realizing this benefit necessitates satisfying the criteria outlined in \cref{sec:fisher}, similar to the specifications employed by the nominal surveys in this analysis. These criteria involve maximising the number density ratio, as well as the difference between the linear biases, quadratic biases, and the products $b_1,b_\Phi$ for each individual tracer.

    \item The advantage exhibited by the multi-tracer approach over a single-tracer analysis, for both power spectrum and bispectrum, could be maintained in a realistic setup, where the accessible scales can be limited. In particular, a bispectrum multi-tracer analysis could improve single-tracer constraints on $\fnll$ up to $\sim50\%$, for an analysis restricted within the very large linear and not shot-noise dominated scales.
    
\end{itemize}

The results presented here do not consider correlations between the various parameters, within the Fisher matrix formalism, but address only the case of the conditional $\fnll$ error. Correlations are important since the power of the multi-tracer approach can be applied to reducing or breaking the various degeneracies between parameters. In order to fully investigate the parametric behaviour of the multi-tracer technique we need to perform a complete Fisher analysis, including the various parameters of the model. This is the subject of the follow-up paper \cite{Karagiannis:2023}.

We considered only the full shape of the power spectrum and bispectrum. We leave the study of the possible gains when using different types of data compression of the bispectrum for follow-up work. Future work will also investigate the inclusion of relativistic and wide-angle effects, which have been shown to be important in the multi-tracer power spectrum and single-tracer bispectrum.

\acknowledgments
We acknowledge an anonymous reviewer for comments that helped us significantly improve the paper. We thank Daisuke Yamauchi for helpful clarifications of their work \cite{Yamauchi:2014ioa}.
DK and RM are supported by the South African Radio Astronomy Observatory and the National Research Foundation (Grant No.\ 75415). JF is supported by Funda\c{c}\~{a}o para a Ci\^{e}ncia e a Tecnologia (FCT) through the
research grants UIDB/04434/2020 and UIDP/04434/2020 and through
the Investigador FCT Contract No.\ 2020.02633.CEECIND/CP1631/CT0002. 
SC acknowledges support from the `Departments of Excellence 2018-2022' Grant (L.\,232/2016) awarded by the Italian Ministry of University and Research (\textsc{mur}) and from the `Ministero degli Affari Esteri della Cooperazione Internazionale (\textsc{maeci}) -- Direzione Generale per la Promozione del Sistema Paese Progetto di Grande Rilevanza ZA18GR02. CC is supported by the UK Science \& Technology Facilities Council Consolidated Grant ST/T000341/1. 
JF, SC and CC thank the University of the Western Cape for supporting a visit to Cape Town, during which this work was initiated. 
This work made use of the South African Centre for High Performance Computing, under the project {\it Cosmology with Radio Telescopes,} ASTRO-0945.

\clearpage

\appendix

\section{Redshift-space kernels and damping factors} \label{app:RSD_kernels}

The general non-Gaussian redshift kernels up to second order for tracer $I$,  neglecting $\order{f_{\rm NL}^2}$ terms, are given by (see e.g.\ \cite{Baldauf2011,Tellarini2016,Karagiannis2018}): 
  \begin{align}
   Z_1^{I}(\bk_i)&=b_1^I+f\,\mu_i^2+\fnl\,\frac{b^I_{\Phi}\,k_i^{\alpha}}{M(k_i)}\;, \label{eq:Z1}\\
   Z_2^{I}(\bk_i,\bk_j)&=b_1^I\,F_2(\bk_i,\bk_j)+f\,\mu_{ij}^2\,G_2(\bk_i,\bk_j)+\frac{b_2^I}{2} +\frac{b_{s}^I}{2}\,S_2(\bk_i,\bk_j) 
   \notag \\ \notag &~~~~
   +\frac{f\,\mu_{ij}\,k_{ij}}{2}\left[\frac{\mu_i}{k_i}\,Z_1^{I}(\bk_j)+\frac{\mu_j}{k_j}\,Z_1^{I}(\bk_i)\right]      +\fnl\,\frac{b^I_{\Psi\delta}}{2}\,\left[\frac{k_i^{\alpha}}{M(k_i)}+\frac{k_j^{\alpha}}{M(k_j)}\right]
   \\  &~~~~
   -\fnl\, b^I_{\Psi}\,N_2(\bk_i,\bk_j)\;.
   \label{eq:Z2}
  \end{align}
Here $f(z)$ is the linear growth rate, $\mu_i=\hat\bk_i\cdot\hat{\bm{z}}$, with $\hat{\bm z}$  the line-of-sight direction, $\mu_{ij}=(\mu_ik_i+\mu_jk_j)/k_{ij}$, and $k_{ij}^2=(\bk_i+\bk_j)^2$.  The kernels $F_2(\bk_i,\bk_j)$ and $G_2(\bk_i,\bk_j)$ are the second-order symmetric SPT kernels (e.g.\ \cite{Bernardeau2002}), while $S_2(\bk_1,\bk_2) = (\hat\bk_1\cdot\hat\bk_2)^2-1/3$ is the  tidal kernel \cite{McDonald2009,Baldauf2012}. The symmetrised $N_2$ kernel is
\begin{equation}
   N_2(\bk_i,\bk_j)=\frac{1}{2}\,\left[\frac{k_i^{\alpha}}{k_j^2\,M(k_i)}+\frac{k_j^{\alpha}}{k_i^2\,M(k_j)}\right]\,\bk_{i}\cdot\bk_{j}\;,
\end{equation}
which encodes the coupling of the PNG potential to the Eulerian-to-Lagrangian displacement field  \cite{Giannantonio2010,Baldauf2011}. The parameter $\alpha=2,1$ for equilateral, orthogonal shapes  \cite{Schmidt2010,Giannantonio2012,Desjacques2016,Cabass:2018roz} and $\alpha=0$ for local shapes \cite{Dalal2008,Slosar2008,Giannantonio2010}.

The non-perturbative RSD fingers-of-god effect, which models the reduction in  clustering power due to nonlinear velocities, is described by a phenomenological model, following \cite{Peacock1994,Ballinger1996}.  Here we use an exponential damping factor, which in Fourier space is a radial convolution, $\delta_{\bk}\rightarrow\delta_{\bk}\exp(-k^2\mu^2\sigma/2)$. Furthermore, in the case of an optical survey, redshift errors smear the galaxy density field along the line-of-sight, which in Fourier space can be described by $\delta_{\bk}\rightarrow\delta_{\bk}\exp(-k^2\mu^2\sigma_r/2)$ \cite{Seo_2003}. The redshift errors propagate into comoving distance errors, where $\sigma_r(z)=c/H(z)\sigma_z(z)$ and $\sigma_z(z)$ is the redshift error of the survey. Following the most general approach, we assume that the damping amplitude is tracer-dependent, so that
  \begin{align}
 {\cal D}_P^{IJ}(\bk)&=\exp\Big[-\frac{1}{2}\,k^2\,\mu^2\,\Big\{ \big(\sigma_P^I\big)^2+\big(\sigma_r^I\big)^2+\big(\sigma_P^J\big)^2+\big(\sigma_r^J\big)^2\Big\}\Big]\;,\\
 {\cal D}_B^{IJK}(\bk_1,\bk_2,\bk_3)&=\exp\Big[-\frac{1}{2}\,\Big\{k_1^2\,\mu_1^2\,\big[\big(\sigma_B^I\big)^2+\big(\sigma_r^I\big)^2\big]+k_2^2\,\mu_2^2 \,\big[\big(\sigma_B^J\big)^2+\big(\sigma_r^J\big)^2\big] \\ \nonumber
 &~~~+k_3^2\,\mu_3^2 \,\big[\big(\sigma_B^K\big)^2+\big(\sigma_r^K\big)^2\big]\Big\}\Big]\;,
  \end{align}
where the redshift dependence has been suppressed for brevity. The amplitude parameters for RSD and redshift errors, for power spectra and bispectra, are independent free parameters. The  fiducial values for RSD are given by the velocity dispersion, i.e.\ $\sigma_P^I(z)=\sigma_B^I(z)=\sigma_v(z)$.

\section{3-tracer case} \label{app:3t}

For three tracers, $I,J,K$ take values $t,t',t''$, and the power spectrum data vector is 
\begin{equation}\label{eq:dvPS_3MT}
\bm{D_P}=\Big[P^{tt},P^{tt'},P^{t't'},P^{t't''},P^{t''t''},P^{tt''}\Big]\;,
\end{equation}
where the expected values of the entries can be derived as in the case of  two tracers, by changing accordingly the tracer-dependent quantities in \cref{eq:PS_MT}.
Likewise for  the bispectrum data vector in the case of three tracers:
\begin{equation}\label{eq:dvBS_3MT}
\bm{D_B}=\Big[B^{ttt},B^{(ttt')},B^{(tt't')},B^{t't't'},B^{(t't't'')},B^{(t't''t'')},B^{t''t''t''},B^{(ttt'')},B^{(tt''t'')},B^{(tt't'')}\Big]\;,
\end{equation}
where the expected values for the first 9 entries can be derived from \crefrange{eq:BS_MT1}{eq:BS_MT2} by exchanging accordingly the tracer indexes in the tracer-dependent quantities. The expected value of the last entry can be defined as
\begin{align} \label{eq:entry_symm_BS}
    \langle \hat{B}^{(tt't'')}\rangle&\equiv B^{(tt't'')}(\bk_1,\bk_2,\bk_3)=\frac{1}{6}\,\Big[\hat{B}^{tt't''}(\bk_1,\bk_2,\bk_3)+\hat{B}^{t''t't}(\bk_1,\bk_2,\bk_3)+\hat{B}^{t't''t}(\bk_1,\bk_2,\bk_3) \nonumber \\ 
    & +\hat{B}^{tt''t'}(\bk_1,\bk_2,\bk_3)+\hat{B}^{t'tt''}(\bk_1,\bk_2,\bk_3)+\hat{B}^{t''tt'}(\bk_1,\bk_2,\bk_3)\Big]\;,
\end{align}
where 
\begin{align}
    \langle \hat{B}^{tt't''}\rangle\equiv\hat{B}^{tt't''}(\bk_1,\bk_2,\bk_3)&={\cal D}_B^{tt't''}(\bk_1,\bk_2,\bk_3)\,\bigg[Z_1^t (\bk_1)\,Z_1^{t'} (\bk_2)\,Z_1^{t''} (\bk_3)\,B_{\rm PNG}(k_1,k_2,k_3) \nonumber \\ 
   &+2\,Z_1^t (\bk_1)\,Z_1^{t'} (\bk_2)\,Z_2^{t''} (\bk_1,\bk_2)\,P(k_1)\,P(k_2)\nonumber \\
   &+2\,Z_1^{t'} (\bk_2)\,Z_1^{t''} (\bk_3)\,Z_2^t (\bk_2,\bk_3)\,P(k_2)\,P(k_3)\nonumber \\
   &+2\,Z_1^t (\bk_1)\,Z_1^{t''} (\bk_3)\,Z_2^{t'} (\bk_3,\bk_1)\,P(k_3)\,P(k_1)\bigg]\;,
\end{align}
and similarly for the other terms, by appropriately permutating the tracer indexes. Note that the symmetrised bispectrum of \cref{eq:entry_symm_BS} has no stochastic terms due to the assumption $P_{\veps\veps_{\delta}}^{IJ}=B^{IJK}_\veps=0$\;.

The data vector of the combined power  spectra and bispectra is defined by combining the data vectors of \cref{eq:dvPS_3MT,,eq:dvBS_3MT}, creating a 16 element array. The covariance of this vector is again a 2x2 block matrix, as in the two-tracer case of \cref{eq:comb_COV}, but now each individual block will have different dimensions. To be more precise:
\begin{itemize}
    \item $\mathrm{Cov}(\bm{P}_i, \bm P_j)$ is a 6$\times$6 matrix,
    \item $\mathrm{Cov}(\bm{B}_i, \bm B_b)$ is a 10$\times$10  matrix,
    \item $\mathrm{Cov}(\bm{P}_i, \bm B_b)$ is a 6$\times$10 matrix.
\end{itemize}

As in the case of two tracers, we consider only the Gaussian part of the covariances, i.e.\ each sub-block is a diagonal matrix. The expression for each sub-block in the $\mathrm{Cov}(\bm{P}_i, \bm P_j)$ block can be easily derived from \cref{eq:PS_cov}, by permuting the tracer indexes accordingly. The same folows for most of the sub-blocks of $\mathrm{Cov}(\bm{B}_i, \bm B_b)$, where all the expressions for the diagonal sub-blocks can be derived from  \cref{eq:blockBB} by interchanging the tracer indices accordingly. The only sub-blocks left to define are those that correspond to the correlations with the symmetrised bispectrum $B^{(tt't'')}$:
\begin{align} \label{eq:blockBB_3TR}
\mathrm{Cov}\l B_a^{ttt}, B_b^{(tt't'')}\r &= \frac{1}{2}\Big( P^{tt}_{a_1}P^{tt'}_{a_2}P^{tt''}_{a_3}+P^{tt}_{a_1}P^{tt''}_{a_2}P^{tt'}_{a_3}\Big) + \text{2\,perm.}\;, \nonumber \\
\mathrm{Cov}\l B_a^{(ttt')}, B_b^{(tt't'')}\r&= \frac{1}{6}\Big[ P^{tt}_{a_1}\Big(P^{tt''}_{a_2}P^{t't'}_{a_3}+P^{t't'}_{a_2}P^{tt''}_{a_3}\Big)+P^{tt}_{a_1}\Big(P^{tt'}_{a_2}P^{t't''}_{a_3}+P^{t't''}_{a_2}P^{tt'}_{a_3}\Big) \nonumber \\
&~~~+2P^{tt'}_{a_1}P^{tt'}_{a_2}P^{tt''}_{a_3}\Big] + \text{2\,perm.}\;, \nonumber \\
\mathrm{Cov}\l B_a^{(tt't'')}, B_b^{(tt't'')}\r &= \frac{1}{6}\Big[ P^{tt'}_{a_1}\Big(P^{tt''}_{a_2}P^{t't''}_{a_3}+P^{t't''}_{a_2}P^{tt''}_{a_3}\Big)+\frac{1}{2} P^{tt}_{a_1} \Big(P^{t't'}_{a_2}P^{t''t''}_{a_3}+P^{t''t''}_{a_2}P^{t't'}_{a_3}\Big) \nonumber \\ &~~~+P^{tt''}_{a_1}P^{tt''}_{a_2}P^{t't'}_{a_3}+P^{tt}_{a_1}P^{t't''}_{a_2}P^{t't''}_{a_3}+P^{tt'}_{a_1}P^{tt'}_{a_2}P^{t''t''}_{a_3}\Big]+ \text{2\,perm.}.
\end{align}
The remaining sub-block covariances can be retrieved by changing accordingly the tracer indices in the above expressions.

The additional expressions needed to derive the power spectra-bispectra cross-covariance in the case of three tracers are:
\begin{align} \label{eq:blockPB_3TR}
\mathrm{Cov}\l P_i^{tt}, B_b^{(tt't'')}\r &= \frac{1}{3}\Big[P^{tt''}(\bk_i)B_{2\delta}^{ttt'}(\bk_{b_1},\bk_{b_2},\bk_{b_3})+P^{tt'}(\bk_i)B_{2\delta}^{ttt''}(\bk_{b_1},\bk_{b_2},\bk_{b_3})\notag \\
&~~~+P^{tt}(\bk_i)B_{1\delta}^{tt't''}(\bk_{b_1},\bk_{b_2},\bk_{b_3})\Big], \nonumber \\
\mathrm{Cov}\l P_i^{tt'}, B_b^{(tt't'')}\r &= \frac{1}{6}\Big[P^{tt'}(\bk_i)B_{2\delta}^{tt't''}(\bk_{b_1},\bk_{b_2},\bk_{b_3})+P^{tt}(\bk_i)B_{1\delta}^{t't't''}(\bk_{b_1},\bk_{b_2},\bk_{b_3})\notag \\
&~~~+P^{tt''}(\bk_i)B_{1\delta}^{tt't'}(\bk_{b_1},\bk_{b_2},\bk_{b_3})+P^{t't'}(\bk_i)B_{1\delta}^{ttt''}(\bk_{b_1},\bk_{b_2},\bk_{b_3})\notag \\
&~~~+P^{t't''}(\bk_i)B_{1\delta}^{ttt'}(\bk_{b_1},\bk_{b_2},\bk_{b_3})
\Big], 
\end{align}
where
\begin{align}
B_{1\delta}^{tt't''}(\bk_{b_1},\bk_{b_2},\bk_{b_3})&=\delta_{ib_1}\Big[B^{tt't''}(\bk_{b_1},\bk_{b_2},\bk_{b_3})+B^{tt''t'}(\bk_{b_1},\bk_{b_2},\bk_{b_3})\Big] \notag \\
&~~~+\delta_{ib_2}\Big[B^{t'tt''}(\bk_{b_1},\bk_{b_2},\bk_{b_3})+B^{t''tt'}(\bk_{b_1},\bk_{b_2},\bk_{b_3})\Big] \notag \\
&~~~+\delta_{ib_3}\Big[B^{t''t't}(\bk_{b_1},\bk_{b_2},\bk_{b_3})+B^{t't''t}(\bk_{b_1},\bk_{b_2},\bk_{b_3})\Big],
\\
B_{2\delta}^{tt't''}(\bk_{b_1},\bk_{b_2},\bk_{b_3})&=\l\delta_{ib_1}+\delta_{ib_2}\r\Big[B^{tt't''}(\bk_{b_1},\bk_{b_2},\bk_{b_3})+B^{t'tt''}(\bk_{b_1},\bk_{b_2},\bk_{b_3})\Big] \notag \\
&~~~+\l\delta_{ib_1}+\delta_{ib_3}\r\Big[B^{tt''t'}(\bk_{b_1},\bk_{b_2},\bk_{b_3})+B^{t't''t}(\bk_{b_1},\bk_{b_2},\bk_{b_3})\Big] \notag \\
&~~~+\l\delta_{ib_2}+\delta_{ib_3}\r\Big[B^{t''t't}(\bk_{b_1},\bk_{b_2},\bk_{b_3})+B^{t''tt'}(\bk_{b_1},\bk_{b_2},\bk_{b_3})\Big].
\end{align}

\clearpage
\bibliographystyle{JHEP}
\bibliography{references}

\end{document}